# Interference suppression techniques for OPM-based MEG: Opportunities and challenges

Robert A. Seymour*, Nicholas Alexander, Stephanie Mellor, George C. O'Neill, Tim M. Tierney, Gareth R. Barnes, Eleanor A. Maguire*

Wellcome Centre for Human Neuroimaging, UCL Queen Square Institute of Neurology, University College London, London WC1N 3AR, UK

***Corresponding authors:**

Robert A. Seymour: rob.seymour@ucl.ac.uk

Eleanor A. Maguire: e.maguire@ucl.ac.uk

**Running title:** Interference suppression in OPM-based MEG

**Highlights:**

- OPM-based MEG faces unique challenges in terms of suppressing interference
- Various hardware solutions are discussed
- Different signal processing strategies are considered
- Example tutorials for reducing interference in OPM-based MEG data are presented

# Abstract

One of the primary technical challenges facing magnetoencephalography (MEG) is that the magnitude of neuromagnetic fields is several orders of magnitude lower than interfering signals. Recently, a new type of sensor has been developed – the optically pumped magnetometer (OPM). These sensors can be placed directly on the scalp and move with the head during participant movement, making them wearable. This opens up a range of exciting experimental and clinical opportunities for OPM-based MEG experiments, including paediatric studies, and the incorporation of naturalistic movements into neuroimaging paradigms. However, OPMs face some unique challenges in terms of interference suppression, especially in situations involving mobile participants, and when OPMs are integrated with electrical equipment required for naturalistic paradigms, such as motion capture systems. Here we briefly review various hardware solutions for OPM interference suppression. We then outline several signal processing strategies aimed at increasing the signal from neuromagnetic sources. These include regression-based strategies, temporal filtering and spatial filtering approaches. The focus is on the practical application of these signal processing algorithms to OPM data. In a similar vein, we include two worked-through experiments using OPM data collected from a whole-head sensor array. These tutorial-style examples illustrate how the steps for suppressing external interference can be implemented, including the associated data and code so that researchers can try the pipelines for themselves. With the popularity of OPM-based MEG rising, there will be an increasing need to deal with interference suppression. We hope this practical paper provides a resource for OPM-based MEG researchers to build upon.

# 1. Magnetoencephalography and interference suppression

Magnetoencephalography (MEG) is a non-invasive neuroimaging technique that measures small magnetic fields outside of the head that originate from current flows throughout the brain (Cohen, 1968). Given its good (~3–5 mm) spatial resolution (Barratt et al., 2018) and excellent (~1 ms) temporal precision, MEG is an increasingly popular tool for cognitive neuroscientists (Baillet, 2017) and clinicians (Hari et al., 2018). Until recently, the only sensors precise enough for performing MEG were superconducting quantum interference devices (SQUIDs). Due to their requirement for superconductivity, these sensors must be housed within a cryogenic dewar. SQUID-MEG systems are therefore stationary, with sensors typically located ~2-3 cm away from the scalp in adults.

One of the primary technical challenges facing MEG data collection and analysis is that the magnitude of neuromagnetic fields measured outside the head are considerably weaker than interfering signals. Neural fields are typically 10-1000 femtotesla (fT, 1 femtotesla = $10^{-15}$ tesla), up to 100,000 times smaller than the Earth's static magnetic field. Additional sources of low-frequency (<20 Hz) interfering noise are common in urban environments, where moving cars, trains and construction work create changes in magnetic field up to 8 orders of magnitude higher than neural signals (Taulu et al., 2014). Narrow-band sinusoidal noise at 50 or 60 Hz is also commonly present in MEG recordings, originating from electrical devices that use alternating current as a power source. Other sources of noise include vibration artefacts from mechanical movement of the MEG device, slight inaccuracies in SQUID design and any irremovable sources of metal on the participant (e.g. dental bridges, cochlear implants).

The primary means of reducing external magnetic interference is to perform MEG recordings inside a magnetically shielded room (MSR), constructed of multiple layers of copper or aluminium and mu-metal, which provide a path for magnetic field lines around the enclosure. For most purposes, two- or three-layer MSRs are sufficient to reduce the remnant field to a few 10's of nanotesla (nT), providing enough shielding for MEG, although greater suppression could theoretically be achieved (Bork et al., 2001). A complementary hardware-based method for reducing interference is to adapt SQUIDs into gradiometers by constructing oppositely wound pick-up coils either oriented axially or on the same plane (Cohen, 1979). Gradiometers effectively suppress uniform fields originating far away from the sensor whilst still picking up nearby sources, for example from the brain. The signal-to-noise ratio (SNR) of gradiometers can be up to 100 times higher than magnetometers (Taulu et al., 2014). For this reason, modern SQUID-MEG systems typically use multichannel arrays of axial/planar gradiometers (e.g. the CTF 275-channel system), or a mixture of magnetometers and gradiometers (e.g. the MEGIN TRIUX™ neo). These hardware-based techniques are complemented by a suite of interference suppression tools that aim to exploit the spatial and temporal properties of MEG data to isolate signals originating from inside the brain whilst suppressing external signals.

*1.1 Optically pumped magnetometers and interference suppression*

Recently, a new generation of MEG sensors called optically pumped magnetometers (OPMs) – also known as atomic magnetometers – have been developed (Boto et al., 2017, 2018; Shah & Wakai, 2013), which overcome many of the limitations of SQUID-MEG systems. OPMs exploit the quantum mechanical properties of alkali atoms to measure small changes in a magnetic field (see Tierney et al., 2019, for a review). The current generation of sensors (e.g. QZFM Gen-2, QuSpin Inc.) has achieved impressive sensitivities of 7-15 fT/Hz from 1-100 Hz. Because OPMs do not require cryogenic cooling, individual sensors can be placed very close to the scalp, resulting in up to five times the sensitivity to cortical sources (albeit with an increased noise floor) compared to conventional SQUID systems (Boto et al., 2016; Iivanainen et al., 2017). Lightweight, whole-head sensor arrays of OPMs are now available (Hill et al., 2020) and are capable of measuring resting-state connectivity with the same robustness as a 275-channel CTF system, but using only 50 sensors (Boto et al., 2021).

A key advantage of OPMs over SQUIDs is that sensors can be placed on the head and fixed in "wearable" arrays within a 3D-printed scanner-cast (see Fig. 1), moving with the head during experimental recordings. Recent neuroscientific work has used OPMs to detect a range of neuromagnetic fields whilst participants made natural head movements, for example, beta-band modulations during a "ping-pong" paradigm (Boto et al., 2018; Holmes et al., 2018), and auditory evoked fields during continuous head movement while standing (Seymour et al., 2021). This opens up a range of exciting experimental and clinical opportunities for OPM-based MEG experiments, including paediatric studies (Feys et al., 2021; Rapaport et al., 2019), and the incorporation of naturalistic movements into neuroimaging paradigms (Roberts et al., 2019; Sonkusare et al., 2019). The wearability of OPMs also means that participants are not required to keep still for long periods of time, aiding participant comfort and potentially improving the quality of experimental data.

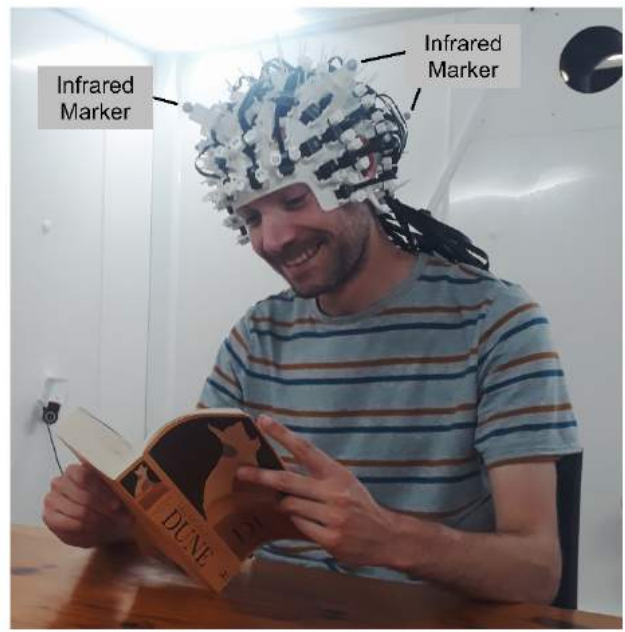

**Fig. 1.** An experimental OPM-based MEG setup. QZFM Gen-2 OPM sensors have been placed in a 3D-printed scanner-cast, with additional custom-made plastic clips to hold the cables in place. Several infrared markers (labelled) have been attached to the scanner-cast for motion capture purposes. The participant is seated in an MSR, performing a naturalistic task (reading), and can move their head freely.

While OPMs undoubtedly have many practical advantages over SQUID systems for MEG, OPMs face some unique challenges in terms of interference suppression. First, the current generation of magnetic sensors based on optical pumping are predominantly magnetometers. As these devices are not superconducting, there is no option (as in SQUID systems) for a flux transformer to couple flux to a single sensor, but rather two or more active elements are required to create a gradiometer (Sheng et al., 2017). This means that optically pumped gradiometer sensors would need to be physically larger than OPMs, and therefore harder to adapt for wearable applications, and would have a slightly higher white noise floor. Second, during OPM experiments where participants move their head, the sensors will also move relative to remnant background magnetic fields within the MSR. This causes very high amplitude, low-frequency artefacts, typically below ~6 Hz (Holmes et al., 2018; Seymour et al., 2021). Without correction, the magnitude of these artefacts easily exceeds any neural signals of interest, and can exceed the dynamic range of current QZFM Gen-2 OPM sensors (~$\pm 5$ nT), resulting in periods of unusable data. Third, more naturalistic paradigms will undoubtedly require OPMs to be integrated with various additional technologies. This may introduce other sources of noise, especially when electrical equipment is required to be located inside the MSR. For example, high-quality motion capture cameras used to track participants' movements introduce narrow-band interference into the data at their operating frequency (e.g. 120 Hz for OptiTrack Flex13 cameras, NaturalPoint Inc.). As another example, Roberts et al. (2019) recently attempted to measure neuromagnetic fields with OPMs whilst participants wore a modified head-mounted display. However, static-field interference from the metallic components in the head-mounted display meant that only signals from the occipital cortex could be successfully measured. Furthermore, high amplitude narrow-band interference meant that OPM data recording was impossible when the sensors were placed too close to the LED screen in the head-mounted display.

It is clear then that, when integrated with technologies within the MSR and/or used for wearable applications, OPM data collection and pre-processing needs to prioritise interference suppression. Here we briefly consider some existing hardware solutions for noise reduction. We then focus on several offline signal processing approaches for interference suppression that researchers might wish to use with their OPM data. The paper builds on the interference suppression literature for SQUID MEG (Taulu et al., 2014) and associated comprehensive guidelines (Gross et al., 2013; Hari et al., 2018), but we hope to highlight OPM-specific issues here. Following theoretical consideration of the pertinent interference suppression methods, two experimental OPM datasets from multi-channel (39-45 sensor) arrays are worked-through for illustrative purposes (Section 5). This includes summary flow diagrams that we hope will enable appreciation of the steps involved in the practical deployment of these techniques, as well as data and analysis code so that researchers can try the pipelines for themselves.

## 2. Hardware for OPM interference suppression

Like SQUID-MEG, OPM-based MEG typically takes place inside an MSR made of copper or aluminium and mu-metal (but see Limes et al., 2020). Current commercially available OPMs operate according to the spin exchange relaxation-free (SERF) principle, and are only capable of measurements in very low background magnetic field levels. By contrast, SQUIDS are stationary, have a higher dynamic range and can operate effectively at any background magnetic field level, as long as it remains consistent over time. Given these additional requirements for OPMs, MSRs for wearable MEG applications should be of the highest quality possible without becoming prohibitively heavy or expensive. For example, a recently developed shielded room designed for OPM-based MEG systems (Magnetic Shields Ltd.) has several design modifications to improve shielding: four half-thickness layers of mu metal, rather than the conventional three, with the grain of two layers turned at 90°; the purity of the mu metal was improved during manufacture; and wider panels of mu metal were produced to create fewer joints. An additional method for ensuring even lower residual fields and gradients in MSRs is via degaussing. This involves applying a sinusoidal current with decreasing amplitude to coils wound around the shielding material inducing magnetic flux in a closed loop (Altarev et al., 2015), resulting in magnetic equilibration of the MSR.

In terms of sensor design, SERF-based OPMs typically possess a set of three inbuilt field cancellation coils to automatically null any residual fields around the vapour cell (Osborne et al., 2018). This is common across most OPMs used for MEG measurements: QZFM Gen-2, QuSpin Inc.; and HEDscan™, Fieldline Inc. (but see Limes et al., (2020) and Kowalczyk et al., (2021) for alternative SERF-free OPM designs). The automated nulling is applied dynamically, but has to be optimised for the sensor's position and orientation at the start of an experimental recording. Where sensors move from their start point through the spatially varying remnant field inside an MSR, the field nulling will become progressively worse. Currently, it is therefore advisable that OPM experiments involving moving participants incorporate frequent breaks into paradigms so that OPMs can be field-zeroed and re-calibrated. The internal coils can theoretically achieve field nulling up to fields of ±50 nT, but in reality, measurement non-linearities are introduced into the data at around ±5 nT. It should be noted that closed-loop OPM systems are developing fast (Fourcault et al., 2021; Kowalczyk et al., 2021; Sheng et al., 2017) and are already commercially available with a dynamic range of 200 nT and bandwidth up to 200 Hz (e.g. HEDscan™, Fieldline Inc.). Technical developments in coil design have also managed to reduce cross-talk between adjacent sensors by a factor of 10 compared with conventional Helmholtz coils (Nardelli et al., 2019). In terms of QZFM Gen-2 operation, it is also possible to dynamically update the internal nulling coils over time based on models of the static field within the MSR (Mellor et al., 2021a). All of these methods (to maintain a zero-field operating point) will increase the dynamic range of the sensors and make OPM systems more robust to low-frequency field drift and movement artefacts that degrade calibration values during open-loop measurements (Iivanainen et al., 2019).

Onboard coils can also be complemented by external biplanar coils (Boto et al., 2018; Holmes et al., 2018) to correct for the remnant background field in the MSR, as measured using reference OPMs placed away from the participant. The compensation field can be dynamically updated for sites in urban environments with time-varying changes in remnant background fields, reducing interference to just ~0.5 nT (Holmes et al., 2019; Iivanainen et al., 2019). Furthermore, detailed field mapping of the MSR can be used to model magnetic field components and gradients in order to update the compensation field produced by the coils. This has recently been shown to further reduce static fields to just 0.29 nT (Rea et al., 2021). However, the use of external custom coils restricts participant movement to a reduced area of the MSR (around 40 cm x 40 cm x 40 cm with current designs). This may not be suitable for experimental setups requiring movement over 40 cm, or for certain cohorts who are likely to exceed this limit over the course of an experiment. To address this issue, Holmes et al. (2021) introduced a novel matrix coil design featuring two 1.6 $m^2$ planes, each containing 24 individually controllable square coils. This resulted in a much larger area of the MSR being nulled, allowing a hyper-scanning two-person ball game to be performed.

## 3. Signal processing strategies for OPMs

External interference cannot be completely removed by the hardware solutions described above, especially in situations where interfering equipment is placed inside the MSR, or where participants are mobile. In the remainder of this article we will turn our attention to signal processing strategies aimed at increasing the SNR of neuromagnetic signals in OPM data. We do not suggest this is an exhaustive list of signal processing strategies, instead we focus on a handful of commonly-used algorithms aimed at addressing various different types of external interference specifically in OPM data.

### *3.1 Regression strategies*

Perhaps the simplest approach to interference suppression is to build up an accurate model of the noise and remove it from the data via regression or other generalised linear approaches.

#### *3.1.1 Reference sensors*

For this purpose, reference sensors can be placed away from the participant to measure interfering fields, but not neural signals. This approach is sometimes referred to as synthetic gradiometry (Fife et al., 1999). At least two OPM reference sensors (assuming one or both are operating in dual-axis mode, see Section 4) should be able to capture the three spatial components of a static noise field. However, for more complex patterns of interference inside the MSR (e.g. during participant movement or time-varying noise fields), a greater number of sensors may be required. Reference channel data can be used to subtract interference from OPM data recorded from the scalp either using simple linear regression or more complex methods such as partial least squares (Adachi et al., 2001) and time-shifted principal component analysis (PCA; de Cheveigné & Simon, 2007). Another technique for optimising synthetic gradiometry is to apply the regression in short over-lapping windows (e.g. 10 s), so as to avoid the detrimental influence of non-stationarities. The utility of synthetic gradiometry is demonstrated in Section 5.2.5.

*3.1.2 Motion capture systems*

In situations where participants are moving freely around the MSR, the movement of the rigid body formed of the head and OPMs can be recorded via a motion capture system. These data can then be used in a multiple linear regression to reduce the magnetic field artefacts covarying with head movement (Holmes et al., 2018; Seymour et al., 2021). A demonstration of this technique is shown in the first example experimental dataset in Section 5. For more complex experimental designs, it may be good practice for researchers to show that the movement data regression reduces the noise floor of OPM recordings to similar levels across experimental conditions and/or participant groups, especially under ~6 Hz.

When motion capture is used for regression, it is important that these data are both low-noise and uninterrupted. Any gaps should be interpolated, and the data should be carefully examined for tracking errors and artefactual spikes. In addition, it is advisable to apply a low-pass filter to the marker trajectories (e.g. 2 Hz bidirectional) before solving the rigid body to ensure that any vibrations or spurious errors in tracking of the motion-capture markers are not introduced into the OPM data. In most instances, a 2 Hz filter would be appropriate to reduce interference during typical head movements. However, if higher frequency movements of the head are required, then the filter could be adjusted, while acknowledging that the trajectories used to solve the rigid body will be noisier. One disadvantage in using motion capture systems is that cameras need to be placed inside the MSR, which can introduce both direct current (DC) and narrow-band noise into the data. Therefore, any benefits of regressing out motion-related artefacts should be balanced alongside the extra noise introduced into the MSR.

Overall, regression-based methods represent a simple, powerful and effective tool for interference suppression. Unfortunately, background magnetic interference is often several orders of magnitude higher than the neuromagnetic signals of interest and highly complex to model. Therefore, when used in isolation, regression-based methods are unlikely to remove all interference from OPM data. It is also worth noting that data from the static OPM reference arrays cannot be used directly to reduce movement-related artefacts through regression. Reference OPMs could, in theory, be fixed to the scanner-cast away from the scalp, similar to the placement of reference sensors on SQUID-MEG systems. However, so far this has proven impractical in terms of scanner-cast design.

## *3.2 Temporal filtering*

One step common to most neuroimaging pre-processing pipelines is the application of temporal filters that aim to attenuate certain frequencies in the data whilst preserving other frequencies of interest. Filtering relies upon the source(s) of interference having a different spectral profile from the neural signals of interest. For detailed discussion of filter theory and design we refer the interested reader to technical articles by Widmann et al. (2015) and de Cheveigné and Nelken (2019). Here, we focus on the practical application of temporal filters to OPM data.

Where OPM data are contaminated by drifts, low-frequency environmental interference or participant movement artefacts, a high-pass filter can be used to increase the SNR. Without correction, sources of high-amplitude, low-frequency interference can severely affect evoked field waveforms and time-frequency spectra when averaged over trials. However, the usual caveats apply when using high-pass filters (Gross et al., 2013), for example, transient changes in a magnetic field may be distorted due to the filtering process (Acunzo et al., 2012), thereby altering the shape and latency of evoked fields (de Cheveigné & Arzounian, 2018; Tanner et al., 2015). Van Driel et al. (2021) also recently showed how high-pass filtering can affect multivariate decoding, by spreading patterns of activity over time, increasing type I errors. Where these issues can be avoided or the experimental question of interest is not latency-related, high-pass filters are an effective way to improve the SNR of both evoked and induced neural activity. In both OPM example tutorials (see Section 5), we use a high-pass filter at 2 Hz to remove low-frequency environmental interference and participant movement artefacts. However, it should be noted that applying a high-pass filter at this frequency precludes the study of low delta-band oscillations, which may be of particular interest in some contexts (e.g. mild traumatic brain injury; Allen et al., 2021). Where interference can be effectively suppressed by other means, OPMs should be capable of measuring delta-band responses (e.g. 0.2-1.5 Hz cortical activity during speech tracking; de Lange et al., 2021).

An offline alternative to classic high-pass filtering is to use detrending, which involves fitting a smooth function to the OPM data (e.g. a low order polynomial) and then subtracting it. Detrending is very sensitive to sensor glitches or railing, and robust implementations should be used for OPM data (as discussed by de Cheveigné and Arzounian, 2018). Where non-stationarities exist in the data, detrending can be made more effective by applying the algorithm in overlapping time windows.

Where high-frequency artefacts are present in OPM data, a low-pass filter can be used. In the first example data analysis tutorial (Section 5.1) we low-pass filter our evoked response data at 40 Hz, but much wider bandwidths have also been used in the OPM literature (Iivanainen et al., 2020). Once again, as with all M/EEG analysis, low-pass filtering can mask wideband noise such as muscle and electrical artefacts in the time domain. OPM researchers may wish to manually inspect their data before applying a low-pass filter.

For narrow-band interference (e.g. from 50/60 Hz line noise or other electrical equipment using alternative current power sources) notch or discrete Fourier transform (DFT) filters are commonly applied to electrophysiological data. In the example data analysis tutorials (Section 5), we make use of an approach termed spectrum interpolation (Leske & Dalal, 2019), which involves transforming the data into the frequency domain, interpolating the noise-contaminated frequency with data from adjacent frequencies, and then transforming the data back into the time domain via an inverse DFT filter. Other options for removing narrow-band interference include Cleanline (Bigdely-Shamlo et al., 2015) which subtracts a sine wave from the data in the time domain estimated adaptively in the frequency domain, and Zapline (de Cheveigné, 2020) which applies a denoising matrix based on spatial filtering.

To conclude on temporal filtering, there are some final points to note that are common to SQUID-based and OPM-based MEG (Gross et al., 2013). First, where possible, filters should be applied to continuous (non-epoched) OPM data to avoid artefacts at the start and end of trials. Data padding at the beginning and end of recordings can also be used to further reduce the influence of edge-artefacts (see https://www.fieldtriptoolbox.org/faq/how_does_the_filter_padding_in_preprocessing_work/). All temporal filter designs are an inherent compromise between an idealised sharp cut-off and the inherent signal distortions in the time domain such as ringing. Researchers might wish to manually inspect their OPM data after temporal filtering for the presence of these time-domain signal distortions. Second, when reporting results, it would be beneficial to comprehensively describe the filter characteristics, including filter type, frequency band, order number and direction. This will allow other researchers to reproduce analyses in full and better evaluate the effects of filter use (de Cheveigné & Nelken, 2019; Pernet et al., 2020).

## 3.3 Spatial filtering

In situations where external interference overlaps in the frequency domain with neural signals of interest, temporal filters cannot be used. Instead, one can take advantage of the multi-channel nature of MEG sensor arrays to separate neural signals from external interference based on their distinct spatial profiles (de Cheveigné & Simon, 2008; Ille et al., 2002; Van Veen et al., 1997). Mathematically this is achieved by applying a spatial filter to the data via a linear algebraic operation, such that:

$$y_n = W_n x_n$$

*Equation 1*

Where the output data *y*, is a weighted sum of the spatial filter *W*, multiplied by the original data *x*, and *n* is the number of independent channels of data.

It is also worth noting that magnetic fields are vectors in 3D space with three independent uniform components and, in quasi-static, source-free space, five independent first derivatives, seven second derivatives, and so on. Therefore, 15 independent sensors would be required to accurately determine an interfering magnetic field up to its second derivative. For this reason, the performance of spatial filtering techniques for interference suppression scales with the number of sensors placed around the head (Taulu et al., 2014).

### 3.3.1 Signal space projection

The first spatial filtering method we will discuss is signal space projection (SSP). This involves projecting experimental MEG data onto a subspace orthogonal to the spatial distribution of interference across channels (Uusitalo & Ilmoniemi, 1997). The subspace is calculated by applying PCA to an empty room MEG recording in order to capture the dominant (those with the largest eigenvalues) spatial patterns related to the environmental interference. It is also possible to compute an interference subspace for physiological artefacts, such as electrocardiogram (ECG) activity, eye-blinks or eye movements, by applying PCA to experimental MEG data epoched around these events. MEG data after SSP will not be linearly independent (i.e. the dimensionality will be decreased by the number of components used for SSP computation). SSP also modifies the statistical properties of the magnetic field vectors originating from the brain, which needs to be accounted for during forward modelling in the course of inverse computations. SSP is a powerful interference suppression technique for SQUID-MEG data, achieving up to 50-60 decibel (dB) reduction for magnetometers based on just two minutes of empty room data (Taulu et al., 2014).

As shown by Tierney et al. (2021a), SSP is also a useful technique for OPM interference suppression. However, SSP relies on the assumption that the properties of the interference are very similar between the empty room recording and the experimental data. This means that a new empty room recording is required every time the spatial configuration of an OPM array changes. In addition, in mobile OPM experiments, the physical location of the sensors will change over time, meaning that the statistical properties of the interference will not match the empty room recording.

*3.3.2 Signal space separation*

One of the most widely adopted spatial interference suppression techniques for MEG is signal space separation (SSS; Taulu & Kajola, 2005). This method relies on the principle that MEG signals can be decomposed, based on Maxwell's equations, into two sets of elementary magnetic fields, one originating from a spherical volume inside the sensor scanner-cast, and another originating from outside. Only signals originating from the internal subspace are retained, thereby reducing external interference from the data. The exact decomposition is based on spherical harmonics expansions that increase in order values from coarse features of the field (low order values) towards increasingly finer details (higher order values). Detailed empirical work has shown that for whole-brain SQUID-MEG data (Taulu & Kajola, 2005), the internal subspace can use an order value of $L_{in} = 8$, and the external subspace $L_{out} = 3$. This corresponds to an SSS basis set of 95 vectors (80 internal, 15 external). For situations where sources of magnetic interference are in close proximity to the sensors (e.g. fixed dental work, implanted stimulators), a temporal extension of SSS can be used (tSSS). This involves computing the temporal correlation between the internal and external subspaces and removing components above some threshold (usually, r=0.99, Taulu & Simola, 2006).

However, SSS/tSSS rely on the assumption of spatial oversampling. When considering whole-head recordings, MEG data are geometrically complex, containing around 100 degrees of freedom (or fewer) that can be separated from external interference (Taulu & Simola, 2006). In modern SQUID-MEG systems, there are many more SQUID sensors than degrees of freedom, thereby satisfying the requirement of spatial oversampling. However, in current OPM-based MEG systems, there are typically 50 sensors or fewer (Hill et al., 2020) due to each sensor's relatively large size and the presence of cross-talk between sensors (Tierney et al., 2019). This means that neuromagnetic fields are not oversampled. In the case of SSS, if the default harmonic expansion values of $L_{in}$ and $L_{out}$ are used with 50 sensors, the solution will not be numerically stable due to high shared variance between the internal and external subspaces. This is especially problematic in the case of tSSS, where thresholds used for SQUID-MEG data (typically r=.90 to r=0.99, with the default implemented in Maxfilter™ set to r=.98) could result in a large proportion of neuromagnetic signals being removed from the data.

To achieve spatial oversampling would require further miniaturisation of OPMs as well as addressing issues of cross-talk between tightly-packed sensors (Tierney et al., 2019, Nardelli et al., 2019). In addition, the advent of OPMs capable of triaxial measurements (Brookes et al., 2021) will increase the effective number of channels on the head, making SSS/tSSS more viable options for OPM data. However, it should also be noted that the stability of the SSS decomposition depends upon the very accurate characterisation of the sensor position, orientation and calibration (Taulu & Kajola, 2005). This would need to be determined for each individual arrangement of OPMs, potentially adding complexity and inaccuracies to SSS denoising, especially in the case of OPMs arranged in non-rigid caps (Feys et al., 2021; Hill et al., 2020) or non-spherical arrays (Tierney et al., 2021b). In summary, SSS/tSSS methods will only be applicable for OPM data when spatial oversampling is achieved and accurate calibration procedures are established.

*3.3.3 Homogeneous field correction*

The lack of spatial oversampling with current OPM-based MEG systems requires the modelling and removal of external interference to be simplified. One approach proposed by Tierney et al. (2021a) involves decomposing the data using only a first order spherical harmonic model (i.e. the interference subspace is set to an order value of $L_{out} = 1$, rather than $L_{out} = 3$ in conventional SSS implementations). This is equivalent to modelling the interference as a spatially constant homogeneous magnetic field (Tierney et al., 2021a). Computationally, this is calculated via the row-wise concatenation of the unit normals representing the sensors' sensitive axes. The modelled interference is then removed from the data via linear regression. As shown in Tierney et al. (2021a), and as we demonstrate in Sections 5.1.6 and 5.2.6, the homogeneous field correction (HFC) approach is simple to apply and substantially reduces external interference in OPM recordings across a broad array of frequencies. Furthermore, because HFC is only reliant on a sensor's orientation and not position, the requirements for the accuracy of sensor calibration will be far less than for standard (i.e. higher order) SSS implementations.

*3.3.4 Independent component analysis*

Independent component analysis (ICA) is a spatial filtering technique used for blind source separation (Makeig et al., 1997; Sejnowski, 1996). It assumes that data are a linear mixture of different neuromagnetic sources over time that are statistically independent, stationary over time and non-gaussian. From this assumption, an un-mixing matrix is estimated based on various statistical properties of the data, including entropy and mutual information (InfoMax). Using example OPM data (see the second example tutorial, Section 5.2.7) we demonstrate the use of the popular fastICA algorithm (Hyvarinen, 1999), which separates components based on directions of maximum kurtosis.

ICA is a widely used interference suppression technique for both EEG and MEG, and is particularly adept at identifying physiological artefacts with distinct spatial topographies, like ECG, eye-blinks, eye movements and muscle activity (Jung et al., 2000). ICA could therefore be useful in mobile OPM experiments where muscle artefacts and naturalistic eye-movements are likely to be present in the data. As with other spatial filtering techniques, the performance of ICA will improve as the number of OPM sensors in whole-head arrays increases, facilitating the statistical separation of sources. It is also worth noting that temporal filters affect the performance of ICA, and that filters applied prior to ICA can differ from those used in the rest of an analysis. For example, the 40 Hz low-pass filter mentioned above might not be suitable prior to ICA as it could remove information required to identify high-frequency components such as muscle artefacts. For mobile neuroimaging, low-frequency artefacts related to movement are likely to affect all channels in a similar way, making them challenging to identify using ICA (Winkler et al., 2015). Therefore, a high-pass filter between 1-2 Hz can improve ICA performance (Klug & Gramann, 2020). The ICA weights produced can then be applied back to the original data where, as discussed above, a 40 Hz low-pass filter may be suitable.

One of the main disadvantages of ICA is the requirement for a manual artefact identification step which can introduce experimenter bias and add significant time to data pre-processing. However, there are automated classification approaches for independent components, for example Corrmap (Viola et al., 2009), IClabel (Pion-Tonachini et al., 2019) and MEGnet (Treacher et al., 2021), which may be adopted as OPM-based MEG systems become more standardised. Reference channel data, ECG and electrooculogram recordings can also be used to guide the automatic removal of components (Hanna et al., 2019). Another disadvantage is that most ICA approaches are probabilistic, meaning that the order of the independent components is arbitrary, and the results may change if re-run. Finally, in relation to OPM-based MEG, one significant disadvantage of ICA is the assumption of stationarity. Where sources of interference move relative to the location of the sensors, common in mobile OPM experiments, standard ICA decompositions are likely to be sub-optimal.

*3.3.5 Source estimation*

So far we have focussed on spatial filtering methods utilising the statistical properties of the data. Now we turn our attention to suppressing interference through source estimation – the process of estimating the actual neural sources of magnetic fields measured outside the head. This relies upon having a forward model describing how neural sources generate magnetic fields at given sensor positions and orientations (Baillet et al., 2001). The reverse operation, going from sensor-space to source-space is an ill-posed problem given that an infinite number of sources could theoretically produce the same sensor-space data. However, by imposing various anatomical and statistical constraints, source estimation algorithms can be used to localise neuromagnetic fields with spatial precision up to several millimetres (Barratt et al., 2018; Boto et al., 2016; Nasiotis et al., 2017). In terms of interference suppression, source reconstruction algorithms help to suppress environmental noise originating from outside the brain whilst increasing signal from neuromagnetic sources.

One popular approach for source estimation is the spatial filtering technique beamforming, which aims to weight MEG data such that signals coming from a particular location of interest are retained whilst all other signals are attenuated (Van Veen et al., 1997). The main assumption behind a beamformer analysis is that no two neuromagnetic sources are correlated over time (Hillebrand & Barnes, 2005). A different set of weights is then calculated sequentially for each location in the brain, usually constrained to a low-resolution cortical mesh or volumetric grid. An important component of the beamformer is that the spatial filter is data-dependent, calculated from the sensor level covariance matrix (Van Veen et al., 1997).

The use of beamforming for interference suppression with SQUID-MEG data is well established (Fatima et al., 2013; Hillebrand & Barnes, 2005; O'Neill et al., 2015). In terms of OPM data, Seymour et al. (2021) recently showed that the use of beamforming constitutes an important step in interference suppression for mobile OPM experiments. Specifically, the spatially correlated low-frequency artefacts from participant movement are suppressed while the dipolar fields from the brain are retained. Beamforming has also been shown to help suppress sources of electrical interference, including high amplitude deep brain stimulation electrodes (Litvak et al., 2010; Oswal et al., 2016). This is promising from an OPM-technology integration perspective for devices that need to be situated inside the MSR.

There are several considerations when using beamformers with OPM data. First, traditional beamformers tend to fail in situations of highly correlated neuronal sources (Van Veen et al., 1997), for example during binaural auditory stimulation (Popov et al., 2018), or cognitive tasks involving bilateral hippocampi (O'Neill et al., 2021). However, there are sparse source reconstruction techniques that are more robust to correlated sources such as "champagne" (Cai et al., 2021; Owen et al., 2012), or "Multiple Sparse Priors" for correlated priors (Friston et al., 2008; López et al., 2014). Second, it should be noted that beamforming benefits from the use of a data-driven covariance matrix (Hillebrand & Barnes, 2005; Woolrich et al., 2011), obtained by epoching and filtering sensor-level data around certain features of interest. In situations where the neural signal(s) of interest are unknown a priori, a broadband covariance matrix, computed over long epochs of data, will negatively impact beamformer performance (Brookes et al., 2008; Hillebrand & Barnes, 2005). Finally, in situations with insufficient OPM data length or bandwidth, beamformer output may be impaired due to an ill-conditioned covariance matrix (Hillebrand & Barnes, 2005; Van Veen et al., 1997). In these cases, a regularisation parameter can be used to improve beamformer performance (Brookes et al., 2008; Woolrich et al., 2011).

Another popular source localisation approach is the $\ell2$ minimum-norm estimate (MNE) (Hamalainen & Ilmoniemi, 1994), that searches for the source distribution with the minimum power ($\ell2$-norm). Implementations of MNE typically incorporate cortical location and orientation constraints (Dale & Sereno, 1993). Further noise normalisation of estimates can be used to create more focal estimates of MEG activity in source space, and for estimates of statistical significance (Dynamic Statistical Parametric Mapping; Dale et al., 2000). Other variations on these approaches have been developed (e.g. sLORETA; Pascual-Marqui, 2002). Unlike beamforming, because these distributed source imaging methods do not use an adaptive spatial filter at each source location, they are theoretically less able to separate out interference from neural activity in very noisy data (Hincapié et al., 2017). While a thorough comparison of different source algorithms for OPM-based MEG is beyond the scope of this article, it would certainly be of benefit to the field.

*3.3.6 Manual removal of artefactual data*
In cases where signal processing strategies are unable to adequately suppress particular sources of interference, a useful approach is to simply remove the artefactual channels and/or data segments via visual inspection. For SQUID-MEG data, it is common to remove bad channels, sporadic high amplitude physiological artefacts, and electronics-related SQUID-jumps (Taulu et al., 2014). This also benefits spatial filtering approaches, which can propagate noise from highly artefactual channels to the rest of the data (e.g. SSS and tSSS; Taulu et al., 2006).

A similar approach can be used with OPM-based MEG data containing high-amplitude interference from idiosyncratic sources, including urban noise (e.g. traffic, trains, construction work), participant movement artefacts and electronics noise. At sites with a more stable interference profile and robust OPM performance, fixed thresholds based on peak-to-peak signal amplitude or Z-scoring could be used instead of manual artefact rejection. Data-driven thresholding and pre-processing tools also now exist for M/EEG data, e.g. Autoreject (Jas et al., 2017). Reiterating general M/EEG guidelines (Gross et al., 2013; Hari et al., 2018), where manual artefacts rejection is employed, we recommend that researchers report the exact criteria used for classifying data as artefactual, as well as reporting the times/channels.

## 4. Multi-axis recordings

Due to the simplicity of OPM sensor design, it is possible to simultaneously measure multiple orientations of the magnetic field (Borna et al., 2020). For example, QZFM Gen-2 sensors can measure two axes of the magnetic field simultaneously (see Section 5.1.2). By splitting the laser-beam within the OPM cell, or using two separate modulation frequencies, triaxial sensors are capable of measuring a full 3D field vector. Detailed simulation work has shown that compared to radial-only oriented sensors, dual-axis and triaxial arrays of OPMs can theoretically measure neuromagnetic fields with greater information content and increased signal amplitude (Iivanaienen et al., 2017). Furthermore, Brookes et al. (2021) recently demonstrated how whole-head arrays of triaxial sensors could substantially improve the spatial filtering properties of a beamformer. This improvement comes about in two separate ways: first, by tripling the channel count, thereby increasing the amount of brain signal measured, and second by reducing the correlation between magnetic field sources, helping to separate neuromagnetic dipolar field shapes from uniform environmental fields originating from outside the brain. Similar improvements in performance are expected for all spatial filtering techniques with triaxial sensors, including SSS (Nurminen et al., 2013) and HFC (Tierney et al., 2021a).

## 5. OPM interference suppression tutorials

In this section, we will outline the practical use of signal processing tools for interference suppression using example OPM data from two separate experiments. Accompanying MATLAB code can be found at https://github.com/FIL-OPMEG/tutorials_interference, which relies upon the *analyse_OPMEG* toolbox(https://neurofractal.github.io/analyse_OPMEG/), custom motion capture processing code (https://github.com/FIL-OPMEG/optitrack) and the Fieldtrip toolbox version 20210606 (Oostenveld et al., 2011). The data presented in these two experiment tutorials is shared openly at https://doi.org/10.5281/zenodo.5539414.

In the first example tutorial, we analyse data from a mobile OPM experiment, in which the participant made natural movements of their head during an auditory evoked field paradigm (Hari, 1990). We use a series of signal processing tools to suppress interference in the data, with a focus on low-frequency artefacts resulting from participant movement. This highlights the unique challenges facing mobile OPM experiments, especially when analysing evoked responses overlapping in the frequency domain (2-40 Hz; Hari, 1990) with movement artefacts (below 6 Hz; Seymour et al., 2021). In the second example tutorial, we focus on time-frequency analysis of beta-band power changes during a finger-tapping task (Cheyne, 2013). For this dataset, the participant remained stationary throughout the recording, and therefore interference at lower frequencies, below ~6 Hz, was much lower than the mobile OPM dataset. However, interference across the frequency spectrum was still high, including for our frequency-band of interest (beta-band: 13-30 Hz). We highlight how signal processing algorithms commonly applied to SQUID-MEG data can also be used on OPM data, with the aim of increasing the robustness and SNR of time-frequency analyses at the sensor-level. In each tutorial, we work through each signal processing step in detail, quantifying its impact on the OPM data. We will finish by summarising the steps taken, and their order, in a flowchart.

Data for the two experiments were collected from the same healthy, right-handed male aged 29 years. He provided written informed consent and the study was approved by the University College London (UCL) Research Ethics Committee.

### *5.1 Measuring auditory evoked fields during participant movement*
Full details of this experiment are reported in Seymour et al. (2021).

#### *5.1.1 Paradigm*
Auditory tones were presented to the participant via PsychoPy (Peirce, 2009) through MEG-compatible ear tubes with Etymotic transducers. The tones had the following characteristics: duration = 70 ms, rise/fall-time = 5 ms, frequency = 500-800 Hz in steps of 50 Hz, inter-stimulus interval = 0.5 s. The volume was adjusted to a comfortable level as specified by the participant. The participant was instructed to stand in the middle of the magnetically shielded room and continually move and rotate their head in any direction they wished. A total of 570 individual auditory tones were presented.

### 5.1.2 *OPM data collection*

OPM data were acquired in a 4-layer MSR (Magnetic Shields Ltd) located at UCL. Forty three OPMs (QZFM Gen-2, QuSpin Inc.) were placed evenly around the head. The sensors were held in place using a participant-specific 3D-printed scanner-cast (Boto et al., 2017), designed by Chalk Studios, using the participant's structural MRI scan. A further two sensors were mounted statically within the room and remote from the participant to act as reference OPMs, however these data were not analysed in this experiment. The OPMs operated in dual axis mode, recording magnetic fields oriented both radial and tangential to the head. Consequently, 86 channels of OPM data were recorded (plus 4 channels of reference OPM data), using a 16-bit precision analog-to-digital converter (National Instruments) with a sample rate of 6000 Hz. In addition, five trigger channels were recorded.

Before the start of the experiment, the MSR was degaussed to minimise the residual magnetic field in the room (Altarev et al., 2015), and the OPM sensors were calibrated and nulled (to minimize static fields using the onboard coils), using a manufacturer-specific procedure.

### 5.1.3 *Head position tracking*

For head position tracking, an array of six OptiTrack Flex13 (NaturalPoint Inc.) motion capture cameras were used. These cameras were placed around the MSR to allow for complete coverage of the head. Six retro-reflective markers were attached to the scanner-cast in multiple fixed positions to form a rigid body. These were tracked passively using the OptiTrack cameras at 120 Hz throughout the experiment. By measuring the joint translation of markers on the rigid body, the motion capture system could calculate the position and rotation of the rigid body while the participant moved within the MSR. More details on the specific steps used for movement data processing can be found in Seymour et al. (2021).

### 5.1.4 *Loading the OPM data and assessing the interference*

The first step in our data processing pipeline involves loading the OPM data, stored in binary format, into MATLAB. This is associated with various descriptor files (e.g. sampling rate, sensor-type, channel name, channel positions) and organised into a data structure compatible with the Fieldtrip toolbox (function: `ft_opm_create`). Of note, there is currently no standard format for OPM data, and therefore this method of loading in the data is customised for the OPM system at UCL. The data are then down-sampled to reduce the sampling rate of the data from 6000 Hz to 1000 Hz, for computational efficiency (function: `ft_resampledata`). Next, we plot the power-spectral density (PSD) of the OPM data using the `ft_opm_psd` function (Fig. 2). An additional line at 15 fT/√Hz is plotted, that corresponds to the field sensitivity value reported by the manufacturer (QuSpin) between 3-100 Hz for these second generation sensors. This allows us to characterise sources of interference greater than this 15 fT/√Hz value, at different frequencies along the spectrum.

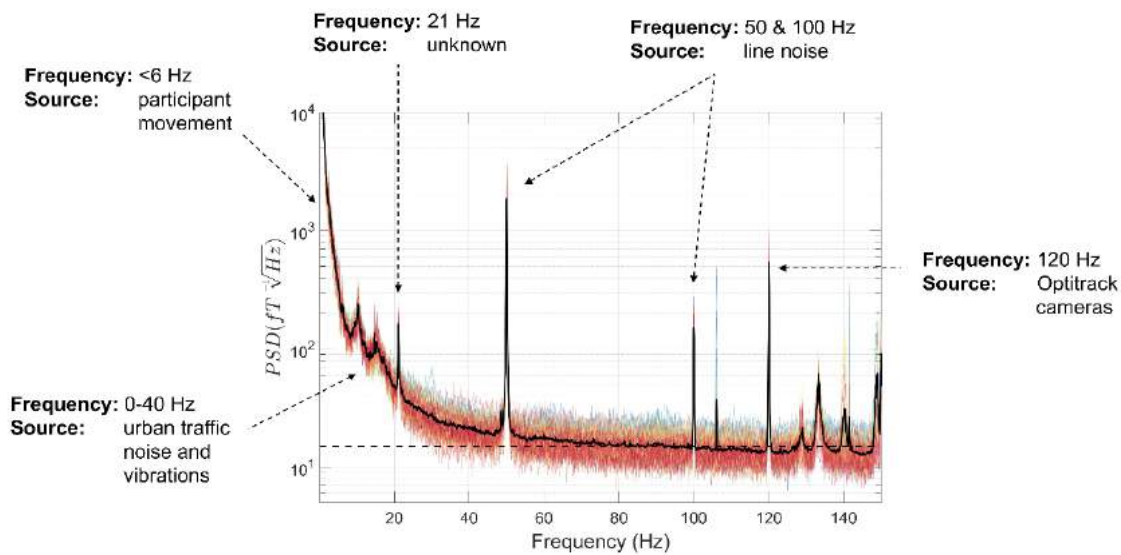

**Fig. 2.** The power-spectral density (PSD) was calculated using 10 s-long windows. Individual channels are plotted in colour, and the mean over all channels is plotted in black. The dotted line on the y-axis corresponds to 15 fT/√Hz. Various sources of interference are labelled in this figure for illustrative purposes.

We can observe various sources of magnetic interference in the data. Below ~6 Hz there are very high PSD values, over $10^4$ fT/√Hz at very low frequencies, resulting from participant movement during the experiment. From 0-40 Hz there are additional sources of low-frequency interference, presumably resulting from motor vehicles, trains and vibrations of the MSR. In addition, there are various narrow-band spikes in the data: at 21 Hz (source unknown), 50 Hz and 100 Hz (line noise), and many more above 100 Hz including at 120 Hz from the LED light source on the OptiTrack cameras.

*5.1.5 Loading head position data and performing a regression*

The position of the rigid body formed of the head, scanner-cast and OPM sensors can be described using three degrees of freedom: right-left, down-up, back-forward (Seymour et al., 2021). A further three degrees of freedom describe the rotation of the rigid body: pitch, yaw and roll (Seymour et al., 2021). These data are stored in a .csv file (`sub-002_ses-001_task-aef_run-003_eul_world.csv`) at a sampling rate of 120 Hz. We load the data into MATLAB using the `csv2mat` function. The movement data are then synchronised with the OPM data (using a trigger sent on channel *FluxZ-A*), and upsampled to 1000 Hz using linear extrapolation to match the OPM data. Fig. 3 shows the head movement data plotted for position (left panel) and rotation (right panel). We can see that the participant did not move for the first 34 s of the recording (presentation of the auditory tones did not start until 34 s). After this, the participant started making various head movements, in all degrees of freedom, continuously until 306 s. The range of these movements exceeded 100cm.

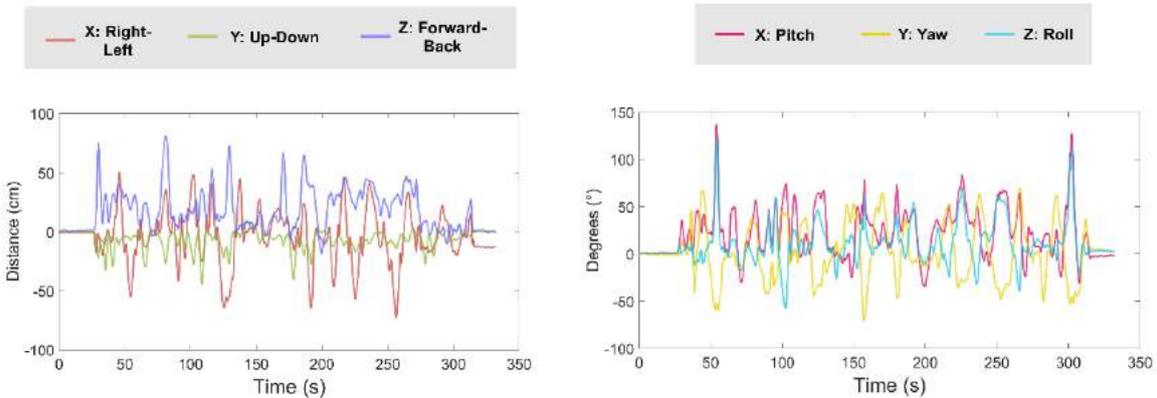

**Fig. 3.** Head movement data plotted over time. Left panel = translations, right panel = rotations. Note the continuous nature of the movements from 34 s to 306 s during the auditory experiment.

The motion capture data were used for interference suppression. Specifically, at each time point a multiple linear regression was performed to reduce the magnetic field artefacts covarying with the motion capture data (Holmes et al., 2018; Seymour et al., 2021; function: `regress_motive_OPMdata`). The regression included the head position data (X, Y, Z) and rotation data (pitch, yaw, roll). Due the presence of non-stationarities in the OPM data, we opted to perform the regression in overlapping 10 second-long windows, sliding from the start to the end of the recording. To measure the amount of interference suppression, we used the formula below, where $PSD_1$ = before interference suppression and $PSD_2$ = after interference suppression.

$$Gain = 20 \times log10 \frac{PSD_1}{PSD_2}$$

*Equation 2*

The results show that the movement data regression step reduced data below 2 Hz by a factor of 30-40 dB, particularly under 0.5 Hz (Fig. 4). However, above 2 Hz, the movement regression step had little impact. This is not surprising given that the raw motion capture data was already low-pass filtered at 2 Hz before further processing (see Section 3.1.2). More generally, this step demonstrates how it is possible to model external interference, in this case by tracking head position during the experiment, and reduce artefacts in the data resulting from the OPMs moving through remnant field gradients in the MSR.

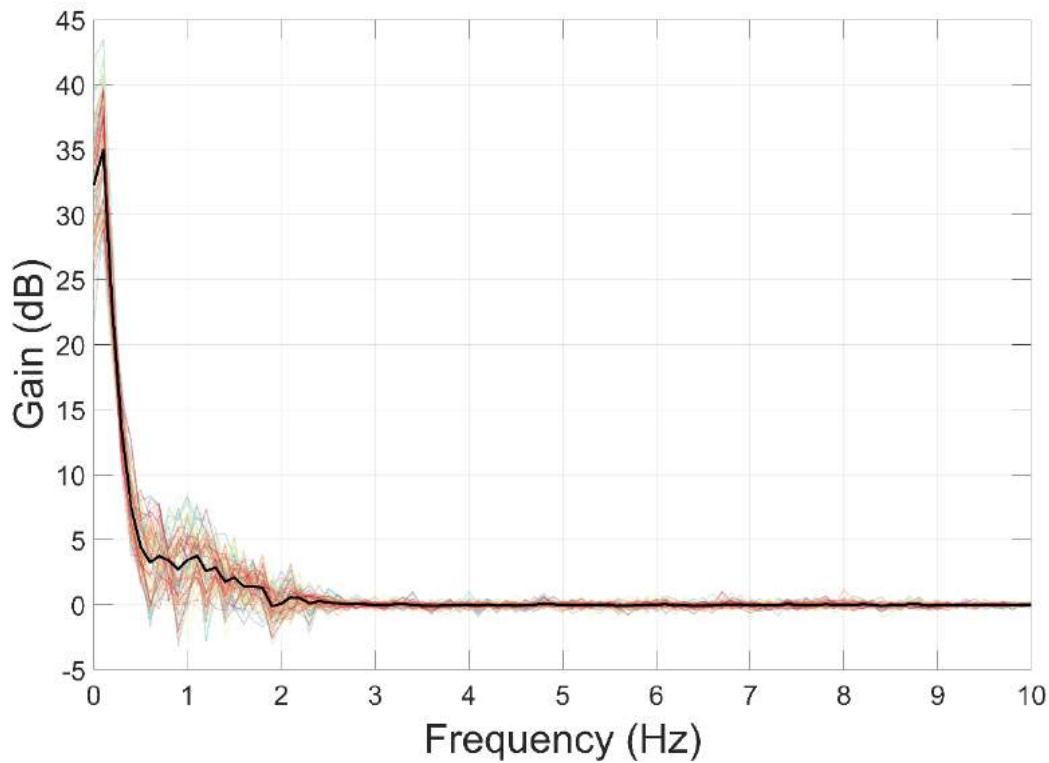

**Fig. 4.** The amount of interference suppression following movement data regression was quantified using Equation 2. Individual channels are plotted in colour, and the mean over all channels is plotted in black.

*5.1.6 Homogenous Field Correction*

Next, we used HFC (Tierney et al., 2021a; function: `ft_denoise_hfc`). This technique approximates magnetic interference as a spatially constant field on a sample-by-sample basis (see Section 3.3.3 and Tierney et al., 2021a for further details). HFC was used in place of more complex interference suppression approaches like SSS (Taulu & Kajola, 2005) or tSSS (Taulu & Simola, 2006), because the OPM data in this example did not satisfy the requirements for spatial oversampling. As discussed in Section 3.2.2, more complex spatial sampling methods will only be easily applicable once OPM systems reach the channel count approaching conventional SQUID-MEG systems (i.e. ~300 channels).

Interference suppression performance following HFC was quantified using Equation 2, where $PSD_1$ = data following movement regression, and $PSD_2$ = data following HFC. Results (Fig. 5) show that HFC reduces interference across a wide range of frequencies, including 0-20 Hz (movement artefacts, traffic, trains, MSR vibrations), the 21 Hz spike of unknown origin, and 50 Hz line noise. This demonstrates how HFC can be used as a broadband interference suppression technique and is appropriate for use with data from an 86-channel array of OPMs.

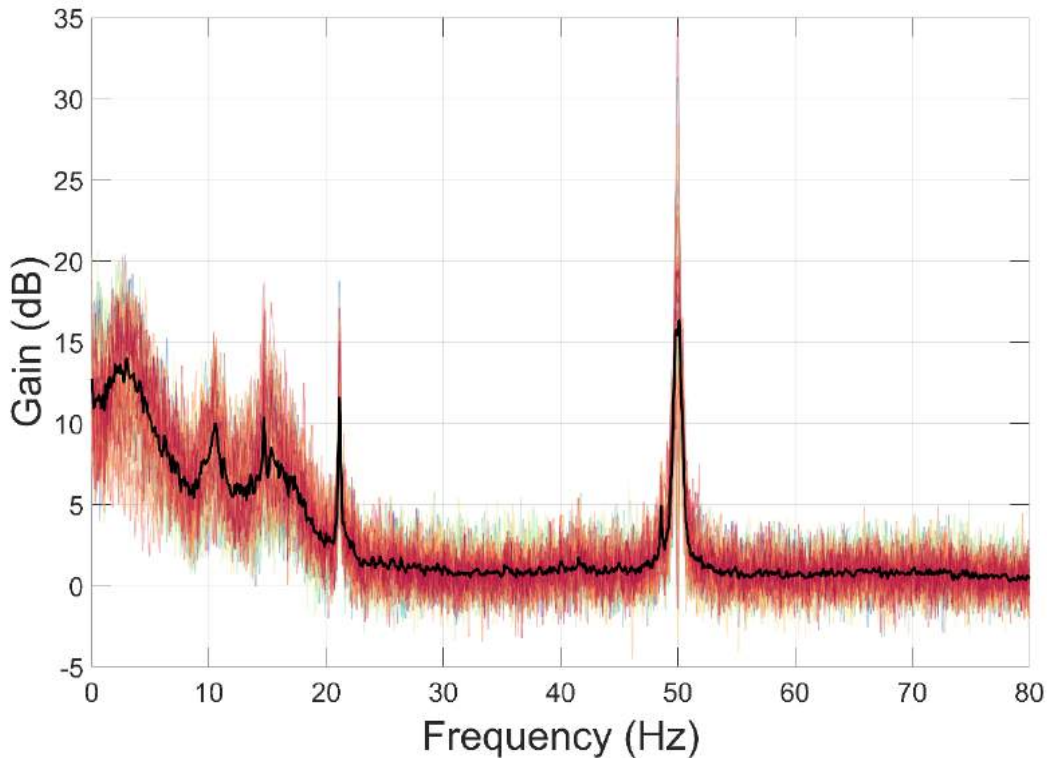

**Fig. 5.** The amount of interference suppression following homogenous field correction (HFC) was quantified using Equation 2. Individual channels are plotted in colour, and the mean over all channels is plotted in black.

## 5.1.7 Bad channel rejection

Plotting the data after HFC, using `ft_databrowser` in 30 s chunks, we identified two bad channels with large idiosyncratic fluctuations and frequent periods of railing: *DS-RAD* and *DS-TAN*. These were removed from the data using `ft_selectdata`.

## 5.1.8 Temporal filtering

The next step of our pipeline utilised temporal filters. For narrow-band sources of interference in MEG data, it is typical to use a notch or DFT filter. However, neither filter is suitable for this dataset because the amplitude of narrow-band interference will change over time as the sensors move in the MSR. Consequently, we adopt a spectral interpolation approach (Leske & Dalal, 2019; also see Section 3.2), using the `ft_preproc_dftfilter` function, with the `cfg.dftreplace = 'neighbour'` option. We used a 1 Hz bandwidth to define the narrow-band interference at 50 Hz, 100 Hz, 106 Hz and 120 Hz, and interpolated using ±1 Hz either side of these frequencies. The PSD was plotted after this step (Fig. 6). Reductions in narrow-band interference can be seen at all four frequencies (shown by yellow arrows).

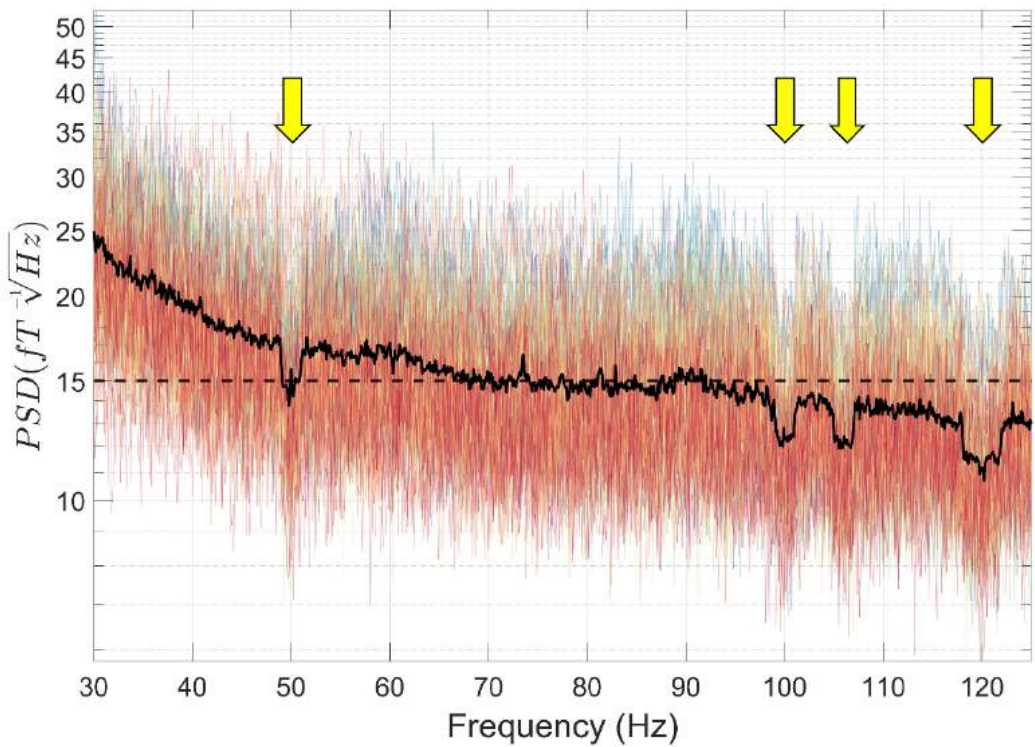

**Fig. 6.** The power-spectral density (PSD) was calculated using 10 s-long windows, following spectral interpolation. Individual channels are plotted in colour, and the mean over all channels

is plotted in black. The dotted line on the y-axis corresponds to 15 fT/ √Hz. The yellow arrows refer to the frequencies that were specified in the spectral interpolation procedure.

This was followed by a high-pass filter at 2 Hz, as implemented in `ft_preprocessing`. Specifically, a 5th order Butterworth filter was used and applied bidirectionally to achieve zero-phase shift. This helped to suppress movement artefacts and other linear trends in the data under 2 Hz, with the aim of increasing the SNR of the auditory evoked response (also see Supplementary Figs. S1-2). At this point we performed a manual artefact rejection step (see Section 5.1.9 below for more details). Finally, a low-pass filter at 40 Hz was applied using a 6th order Butterworth filter applied bidirectionally, as implemented in `ft_preprocessing`. This step also had the effect of reducing high-frequency interference above 40 Hz (Fig. 7 and Supplementary Figs. S3-4). Note that we applied the filters on the continuous data so as to avoid edge artefacts in the time-domain.

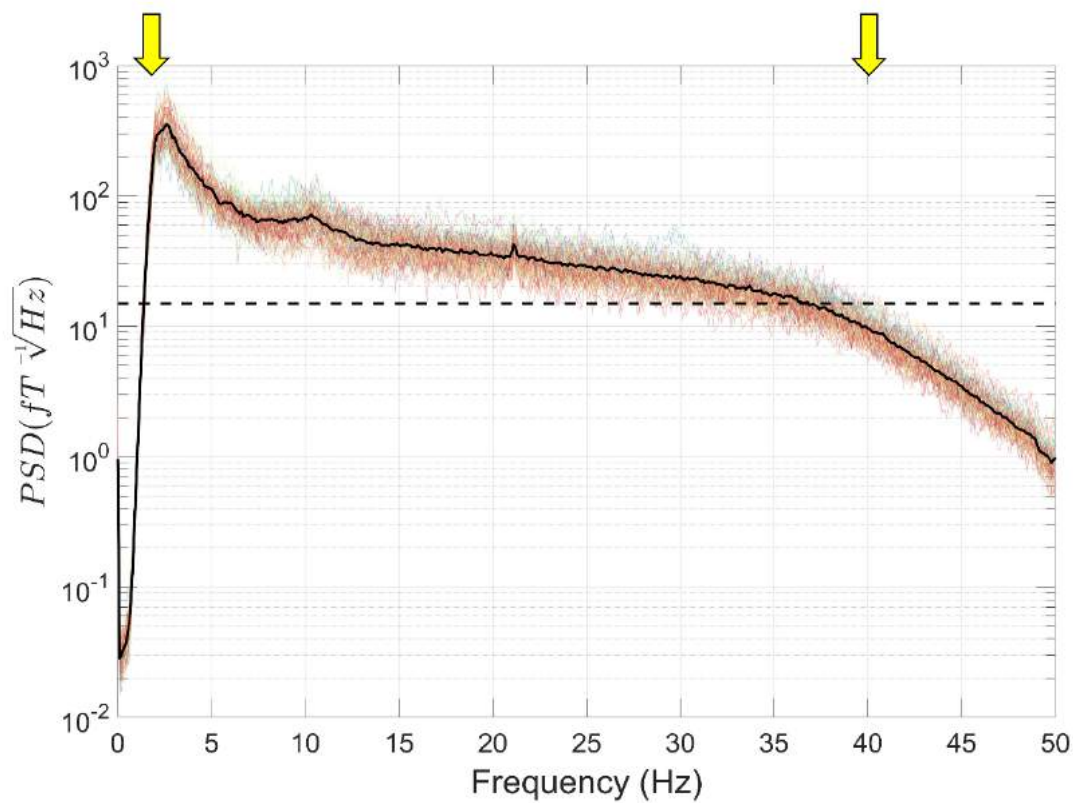

**Fig. 7.** The power-spectral density (PSD) was calculated using 10 s-long windows, following temporal filtering. Individual channels are plotted in colour, and the mean over all channels is plotted in black. The dotted line on the y-axis corresponds to 15fT/ √Hz. The yellow arrows specify the frequencies at which the high (2 Hz) and low-pass (40 Hz) filters have been applied.

*5.1.9   Manual artefact rejection*

We undertook a manual artefact rejection step to identify segments of the data still contaminated by interference. The pre-processed data were visualised in 10 s chunks using the interactive functionality of `ft_databrowser`, before low-pass filtering, so that high-frequency artefacts could be more easily identified. We focussed on identifying transient (less than 100 ms) and very large shifts (<5 picotesla, pT) in the OPM data that appeared on all channels.

In total, we marked 10.6 s of the 352 s data recorded as artefactual, the exact time indices of which are specified at https://github.com/FIL-OPMEG/tutorials_interference (`arft.mat`). This step allowed us to identify and remove artefacts from the data which would have otherwise reduced the SNR of the auditory evoked field. During manual inspection of the data we could not identify either eye-blink or cardiac artefacts, and therefore independent components analysis was not used for this dataset.

*5.1.10 The gradual removal of interference*

To demonstrate how the previous pre-processing steps sequentially removed low-frequency interfering magnetic fields from the data, we calculated the maximum change in field over 1 s chunks of the continuous data. As seen in Fig. 8A, compared with the raw data (red line, $10^2$-$10^3$ pT change per 1 s chunk), each subsequent pre-processing step reduced the variation in magnetic fields to around 1-10 pT per 1 s chunk following temporal filtering (black line).

In the frequency domain, we calculated mean PSD values across all MEG channels after each subsequent pre-processing step. As seen in Figure 8B (left panel), at low frequencies (0-5 Hz) movement regression reduced interference below 2 Hz and HFC from 0-5 Hz. This was followed by a high-pass filter at 2 Hz (see black line). For higher frequencies (Figure 8B, right panel), HFC reduced PSD values across the frequency spectrum, and the temporal filtering step successfully suppressed frequencies above 40 Hz.

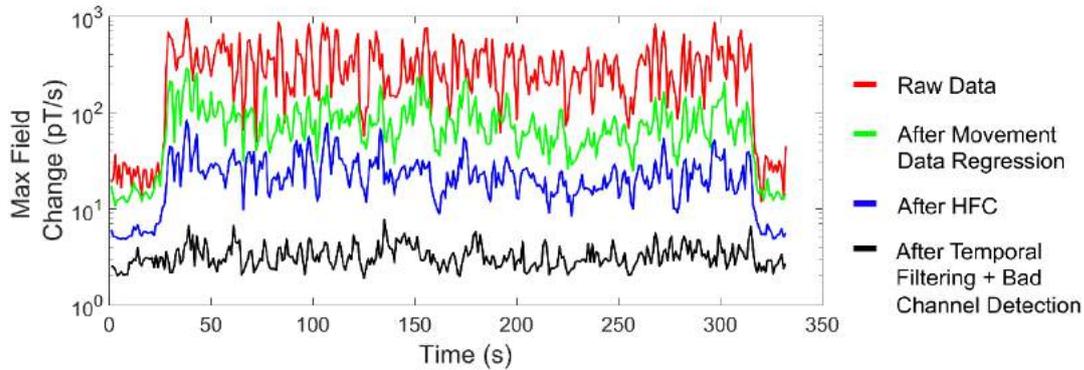

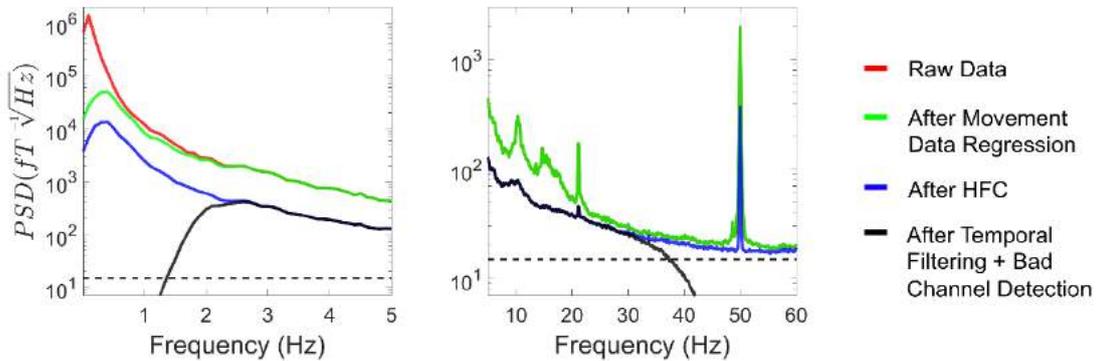

**Fig. 8.** (A) The maximum magnetic field change was calculated over 1 s chunks as each pre-processing stage was applied sequentially for the raw data, after movement data regression, after homogenous field correction (HFC) and after temporal filtering. (B) The power-spectral density (PSD) was calculated using 10 s-long windows, and averaged over channels, for the raw data and after each subsequent pre-processing step. The dotted line on the y-axis corresponds to 15 fT/√Hz.

### 5.1.11 Sensor-level auditory evoked fields

Pre-processed data were epoched into trials of 0.7 s (-0.2 s pre-stimulus, 0.5 s post-stimulus onset), using triggers sent at the onset of auditory tone presentation (OPM channel *NI-TRIG*). Any trial that overlapped with data marked as artefactual in the previous manual inspection step was removed. This resulted in the removal of 20 trials out of a total of 570. The remaining data were averaged and baseline corrected using the 0.1 s of data before stimulus onset. A one sample student t-test (compared to a null of zero) was conducted at each time point across trials, and event-related activity was plotted for each sensor. Results (Fig. 9, upper panel), show the presence of an auditory evoked potential around 100 ms corresponding to the classic M100 (Hari, 1990; Taulu & Hari, 2009). A fieldmap was produced to demonstrate the topography of the M100 evoked response (80-120 ms post stimulus onset) for sensors oriented radially to the head (the tangential components being more difficult to visualise).

For illustrative purposes, we repeated this procedure for the raw data (Fig. 9, bottom panel). T-values were close to 0, with no clear evoked waveform on any channel. By comparing t-values before and after pre-processing, the effective sensor level SNR increase at the peak of the M100 evoked response was 21.37 dB.

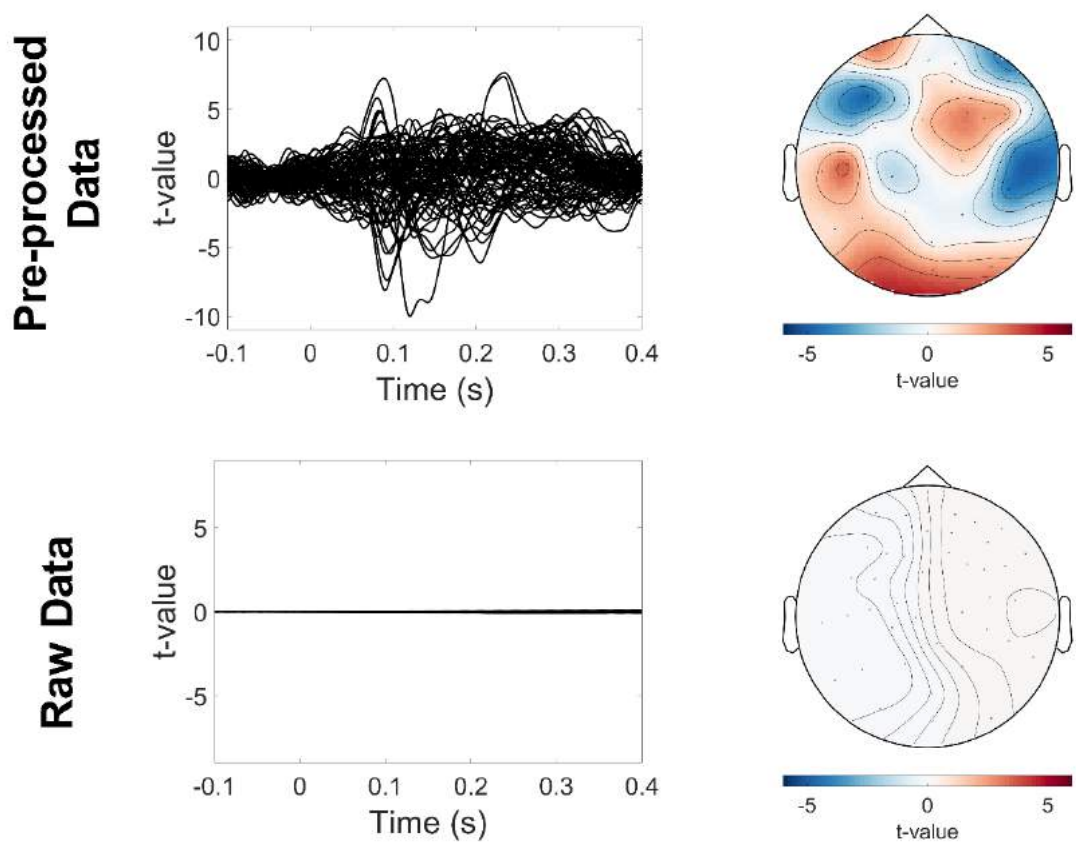

**Fig. 9.** Sensor-level evoked fields. For the pre-processed data (upper panel) and the raw data (lower panel), evoked waveforms for each channel were plotted (left panels). A 2D fieldmap (right panels) was also produced for evoked data from 0.08 s to 0.12 s post-stimulus onset. The fieldmap only shows magnetic fields oriented radially to the head (the tangential components being more difficult to visualise).

*5.1.12 Beamforming to an auditory cortex region of interest*

Having established that our sensor-level pre-processing pipeline helped to increase the SNR of the M100 auditory evoked field, we next used a region of interest (ROI) beamforming approach to suppress the interference even further.

For computation of the forward model, the participant's T1-weighted structural MRI scan was used to create a single-shell description of the inner surface of the skull (Nolte, 2003). We defined a ROI in right primary auditory cortex. Using `ft_volumenormalise`, a nonlinear spatial normalisation procedure was used to warp Montreal Neurological Institute (MNI) coordinates [-48 -22 4; 48 -22 4] from the canonical MNI brain to the participant's MRI scan. These MNI coordinates overlap with the location of bilateral primary auditory cortex from where auditory evoked fields are known to arise (Hari, 1990; Kowalczyk et al., 2021). Source analysis was conducted using a linearly constrained minimum variance (LCMV) beamformer (Van Veen et al., 1997), using the function `ft_sourceanalysis`.

Due to the highly correlated near simultaneous neural activity in bilateral auditory regions evoked by auditory stimulation, traditional beamformers often yield suboptimal results for auditory data (Brookes et al., 2007; Sekihara et al., 2002; Van Veen et al., 1997). Consequently, we opted to construct a dual source model, in which the beamformer is simultaneously computed on dipoles in the left and right auditory cortex (Popov et al., 2018). Based on recommendations for optimising MEG beamforming (Brookes et al., 2008), a regularisation parameter of lambda = 0.1% was used. Beamformer weights were calculated by combining lead-field information with a sensor-level covariance matrix computed from the unaveraged single-trial data from 0-0.5 s post-stimulus onset using function `ft_timelockanalysis`. The spatial filter was then right-multiplied with the pre-processed sensor-level data to obtain an A1 virtual channel. These data were averaged, using `ft_timelockanalysis`, and a one sample student t-test was conducted at each time point across trials. The results (Fig. 10) show a clear evoked waveform at around 100ms corresponding to the auditory M100 response (Hari, 1990; Kowalczyk et al., 2021).

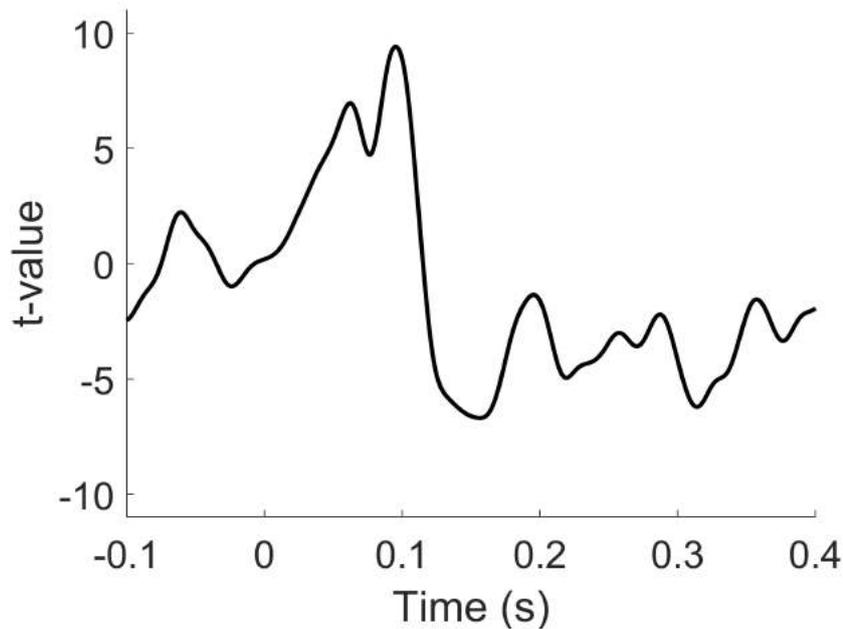

**Fig. 10.** Evoked waveform plotted from the auditory cortex ROI following beamforming. Note the increase in t-values at ~0.1 s, corresponding to the auditory M100 evoked field.

*5.1.13 Summary of the first tutorial*

In this tutorial, we analysed OPM data from a participant standing up and constantly moving their head during an auditory evoked field paradigm. The data contained very high amplitude low frequency artefacts from the sensors moving through remnant background magnetic field gradients in the MSR. Despite performing the experiment inside a degaussed MSR (Altarev et al., 2015), the amplitude of these artefacts was far larger than the neuromagnetic signal of interest (AEFs in this subject were ~220 fT), and further signal processing was required. This is likely to be the case for any OPM experiments involving natural participant movement, even with external nulling coils (Rea et al., 2021). The data were also contaminated by other sources of interference across the frequency spectrum (see Fig. 2). Fig. 11 shows the interference suppression pipeline used in this tutorial. We focussed first on attenuating low-frequency movement artefacts by regressing motion capture data from the OPM data. HFC was then used to reduce interference across the frequency spectrum. Temporal filters were used to filter the data between 2-40 Hz, as is commonly performed in evoked magnetic field analysis (Hari, 1990; Taulu & Hari, 2009). Before the low-pass filtering, manual artefact rejection was performed to remove trials containing spontaneous high-frequency interference. At the sensor-level, the combination of these steps allowed us to measure the M100 auditory evoked field (we were unable to observe the M100 when averaging the raw data). As a final step, we used a ROI-based beamforming approach to reduce interference even further, and characterise the auditory evoked field response in source-space (Seymour et al., 2021). Overall, this example tutorial has demonstrated how, despite large movement-related artefacts alongside a variety of interference sources across the frequency spectrum, signal processing techniques can be used to successfully measure neuromagnetic evoked field data during participant movement.

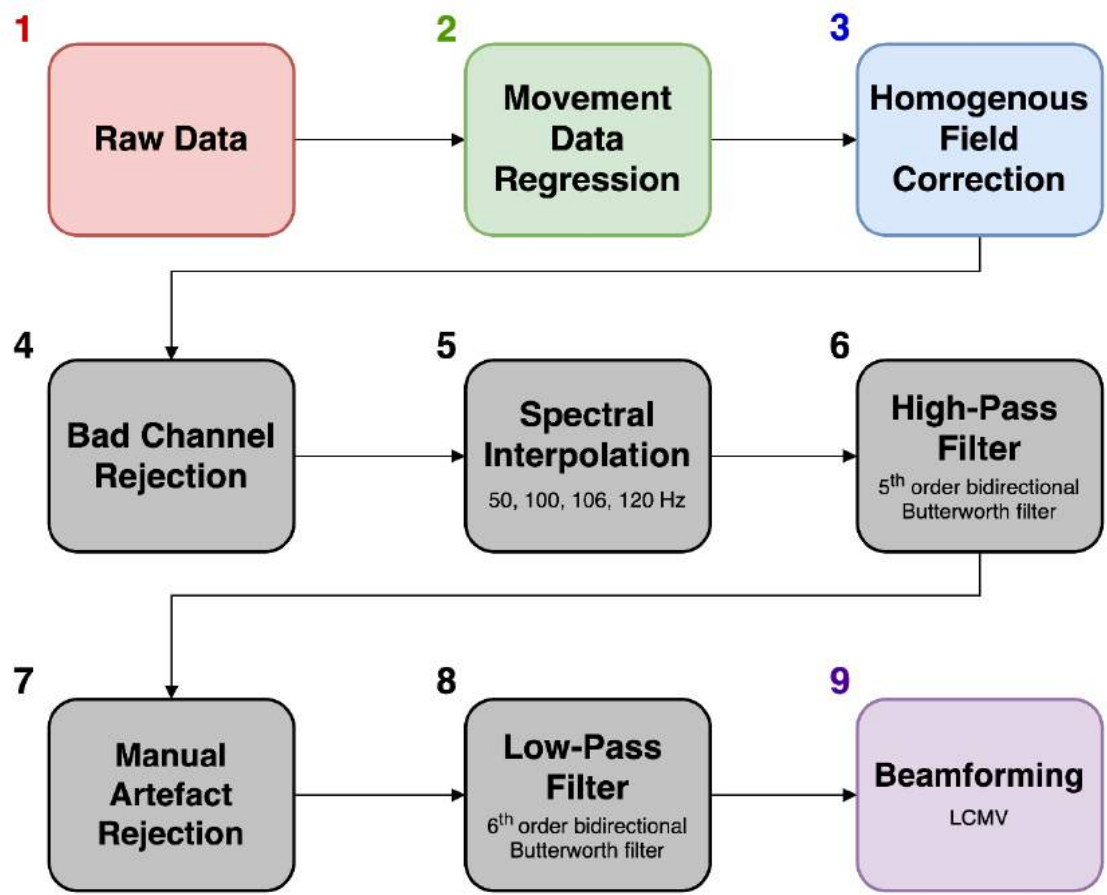

**Fig. 11.** Flow diagram demonstrating the order of the interference suppression steps taken in the first example tutorial. Note that colour coding aligns with the steps outlined in Fig. 8.

*5.2 Measuring beta-band (13-30 Hz) power changes during finger-tapping*

In the second example tutorial, an OPM dataset is analysed in which a participant performed a finger-tapping task. We focus on characterising power changes within a specific frequency band (beta-band, 13-30 Hz), using time-frequency analyses at the sensor-level. Unlike the first tutorial, the participant kept their head still during the experiment, and therefore low-frequency interference was lower. The steps outlined in the tutorial will be similar to standard pre-processing pipelines used for conventional SQUID-MEG analysis (Gross et al., 2013).

However, because the noise floor of current generation OPMs (e.g. QZFM Gen-2) is slightly higher than SQUID-based gradiometer systems (e.g. CTF 275), and the data were acquired in a noisy urban environment (Central London, UK), the successful application of interference suppression algorithms is critical, especially for sensor-level analysis.

*5.2.1  Paradigm*

The experiment was conducted in a 4-layer MSR with the participant sitting on a plastic chair facing a screen. An image was projected onto the screen through a wave-guide using a projector placed outside the MSR. Each trial started with the presentation of a fixation cross (white on black background) for 7-8 s (randomly jittered across trials). When the fixation cross changed colour from white to red, the participant was instructed to lift up their right index finger and perform a rapid tapping motion in the air for 2.5 s, until the fixation cross changed colour back to white. This was repeated for a total of 100 trials. The participant was instructed to remain seated and to keep as still as possible during the recording.

*5.2.2  OPM data collection*

Data collection in this experiment was similar to that for the first experiment, as outlined in Section 5.1.2. The only difference was that 39 OPMs were placed around the head. A further two sensors were located away from the participant to act as reference OPMs, and in this experiment we used these reference OPMs for synthetic gradiometry (Fife et al., 1999). In total 78 channels of OP-MEG data were recorded (plus 4 channels of reference OPM data). Five trigger channels were also recorded.

*5.2.3  Loading the OPM data and assessing the interference*

The OPM data were loaded into MATLAB using the function `ft_opm_create`, and then down-sampled from 6000 Hz sampling rate to 1000 Hz using `ft_resampledata`. Next, the PSD of the OPM data was plotted (Fig. 12) using `ft_opm_psd`, with an additional line at 15 fT/√Hz corresponding to a sensor's field sensitivity reported by the manufacturer. Compared with the data shown in the first tutorial, there is much lower interference below ~6 Hz, because the participant was sitting down and instructed to keep still during the experiment. However, between 0-40 Hz, PSD values are still greater than 15 fT/√Hz, presumably resulting from urban environmental interference alongside vibrations of the MSR. In addition, there are various narrow-band PSD spikes in the data at 21 Hz, 41.5 Hz, 83 Hz (sources unknown), 50 Hz and 100 Hz (line noise). There are also many more PSD spikes above 100 Hz. However, as finger-tapping mainly modulates lower frequencies, for example mu (8-13 Hz) and beta (13-30 Hz) rhythms (Barratt et al., 2018; Cheyne, 2013; Rosenbaum, 2009), we ignore these higher frequency spikes.

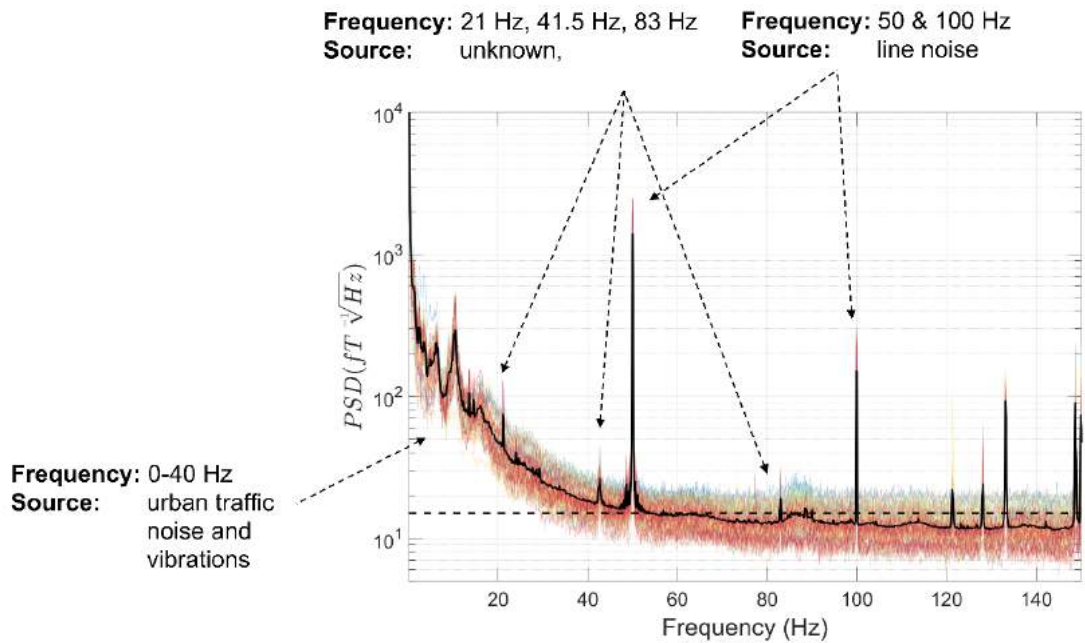

**Fig. 12.** The power-spectral density (PSD) for the second experiment was calculated using 10 s-long windows. Individual channels are plotted in colour, and the mean over all channels is plotted in black. The dotted line on the y-axis corresponds to 15 fT/√Hz. Various sources of interference are labelled in this figure for illustrative purposes.

*5.2.4 Temporal filtering*

As a first step to interference suppression, the data were temporally filtered. To remove high amplitude narrow-band interference we used a spectral interpolation approach (Leske & Dalal, 2019; also see Section 3.2). For this, we used a 1 Hz bandwidth to define the narrow-band interference at 21 Hz, 83 Hz and 100 Hz, and interpolated using +-1 Hz either side of these frequencies. This step was performed at this stage because high amplitude narrow-band interference can substantially reduce the effectiveness of synthetic gradiometry. We also applied a high-pass filter at 2 Hz (5th order Butterworth filter applied bidirectionally to achieve zero-phase shift), to help suppress artefacts resulting from small involuntary movements of the head (the OPM sensors being unconstrained and attached to the head) and other linear trends from the data under 2 Hz. At this stage, we performed a manual data inspection step (in 10 s chunks) using the interactive functionality of `ft_databrowser` to investigate whether the data contained any periods of high-frequency interference, similar to the data in the first tutorial. No such interference was observed. This was followed by a low-pass filter applied at 80 Hz using a 6th order Butterworth filter applied bidirectionally. Note that all filters were applied on the continuous data so as to avoid edge artefacts in the time domain.

Unlike the first example tutorial, temporal filters were applied as the first pre-processing step in our pipeline. This is because the overall amplitude of magnetic field change throughout the experiment was far lower as a result of the participant keeping their head stationary. This reduces the risk of temporal filtering artefacts like ringing or the introduction of filter response peaks into the data (de Cheveigné & Nelken, 2019). Nevertheless, we performed a manual data inspection step (in 10 s chunks) after temporal filtering using the interactive functionality of `ft_databrowser` in order to check for large sinusoidal changes in amplitude characteristic of filter ringing. No evidence of such artefacts was found.

*5.2.5 Synthetic gradiometry*

Next, the two OPM reference sensors (*N0* and *N4*), which were mounted statically and remote from the head, were used to subtract interference from the OPM sensors located on the head via a simple linear regression (Fife et al., 1999). This synthetic gradiometry was implemented using the function `ft_opm_synth_gradiometer_window`. Before applying the regression, the reference data were low-pass filtered at 20 Hz and high-pass filtered at 20 Hz (6th order Butterworth applied bidirectionally) to separate the data into two frequency bands, 2-20 Hz and 20-80 Hz. This was based on the separability between lower frequency interference (urban environmental noise and MSR vibrations), and higher frequency interference (50 Hz line noise and other narrow-band interference). In addition, due to the presence of non-stationarities in the OPM data from low-frequency environmental noise, the regression was applied using 100 s overlapping chunks of data.

The performance of synthetic gradiometry for interference suppression was quantified using Equation 2, where $PSD_1$ = data following temporal filtering between 0-80 Hz, and $PSD_2$ = data following synthetic gradiometry. Interference was reduced mainly between 8-14 Hz and for 50 Hz line noise (Fig. 13). Note that synthetic gradiometry did not reduce the PSD spikes at 21 Hz and 41.5 Hz, suggesting that the two reference sensors did not measure this particular source of noise, or that more variance could be explained by suppressing orthogonal noise sources (such as the 50 Hz). Similarly, despite the high PSD values below ~8 Hz (see Fig. 12), no reduction in interference was observed below 8 Hz following synthetic gradiometry. These low-frequency artefacts likely result from small involuntary head movements over the course of the recording, which cause the sensors to move through remnant magnetic field gradients inside the MSR. Note that although the participant was instructed to sit down and keep still, the head, scanner-cast and OPM sensors were all unconstrained and could have easily moved a small amount over the course of the experiment.

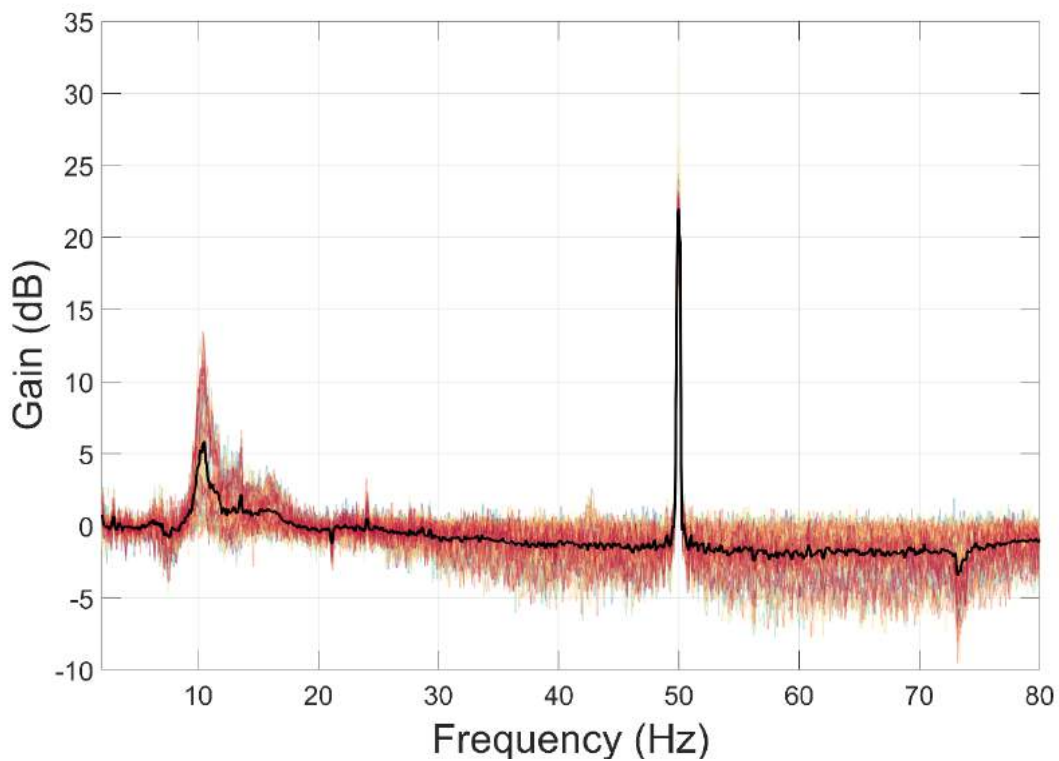

**Fig. 13.** The amount of interference suppression afforded by synthetic gradiometry was quantified using Equation 2. Individual channels are plotted in colour, and the mean over all channels is plotted in black.

*5.2.6 Homogenous Field Correction*

As in the first experiment (Section 5.1.6), we next used HFC (Tierney et al., 2021a; function: `ft_denoise_hfc`). Interference suppression performance following HFC was quantified using Equation 2, where $PSD_1$ = data following synthetic gradiometry, and $PSD_2$ = data following HFC. As in the first tutorial, HFC reduced interference across a wide range of frequencies including 0-20 Hz (movement artefacts, urban noise, MSR vibrations), the 21 Hz and 41.5 Hz spikes of unknown origin, and 50 Hz line noise (Fig. 14).

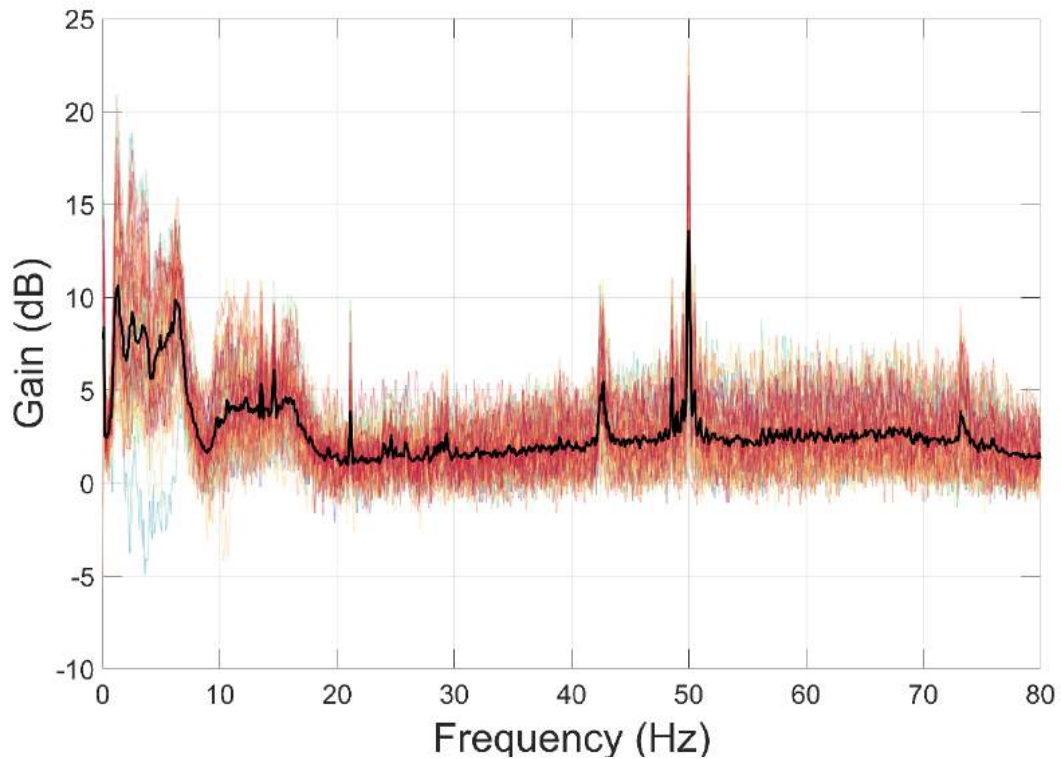

**Fig. 14.** The amount of interference suppression following HFC was quantified using Equation 2. Individual channels are plotted in colour, and the mean over all channels is plotted in black.

*5.2.7   ICA*

The final pre-processing step was the application of ICA to identify and remove magnetic artefacts from non-neural physiological sources (Fatima et al., 2013; Makeig et al., 1997). The popular fastICA algorithm was employed to decompose the data into independent components using `ft_componentanalysis`. For computational efficiency, we specified that the function return 50 components; physiological artefacts are typically returned within the first few components. In addition, a random seed was used (`cfg.randomseed = 454`) so that the ICA decomposition could be reproduced. The PSD, topography and time-course of each independent component was manually inspected. Component 6 is likely to correspond to eye-blink and/or eye movement artefacts (Fig. 15, upper panel). Its PSD is dominated by low-frequency activity with no clear alpha-band peak, its topography is dominated by power close to the location of the eyes, and its time-course corresponds to large, sporadic dipolar changes in magnetic field lasting approximately 0.2-0.5 s. Component 10 is likely to correspond to ECG activity (Fig. 15, lower panel). Its PSD is dominated by low-frequency activity, with no clear alpha-band peak, its topography shows maximal power for sensors at the side of the head (closest to the heart), and its time-course shows regular peaks typical of ECG activity. These two components were removed from the data using `ft_rejectcomponent`.

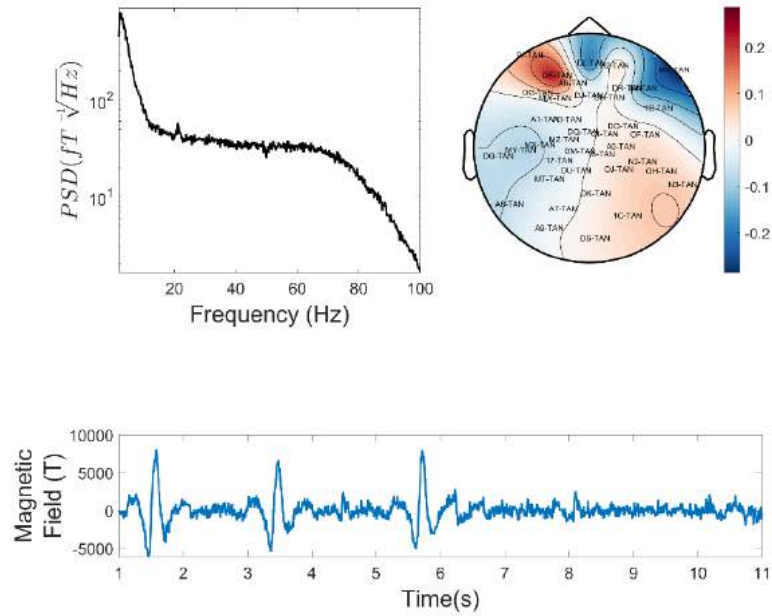

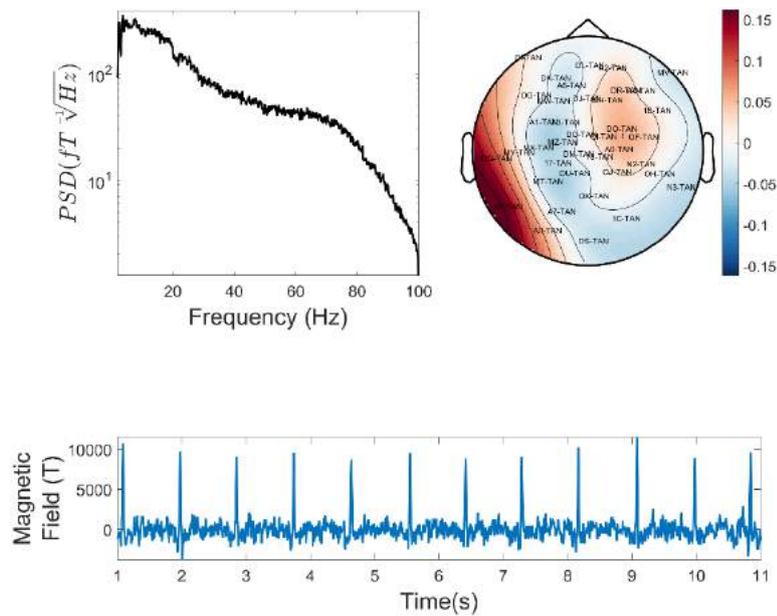

**Fig. 15.** The power spectral density (PSD), topography and representative time-course was plotted for ICA Component 6 (upper panel) and Component 10 (lower panel). These components appear to capture predominantly non-neural physiological interference. Note that only sensors oriented radially to the head were used to construct the field-map, for easier interpretation of the topographies.

Equation 2 was used to compare the PSD of OPM data before and after ICA. As shown in Fig. 16, ICA has predominantly removed interference below ~5 Hz.

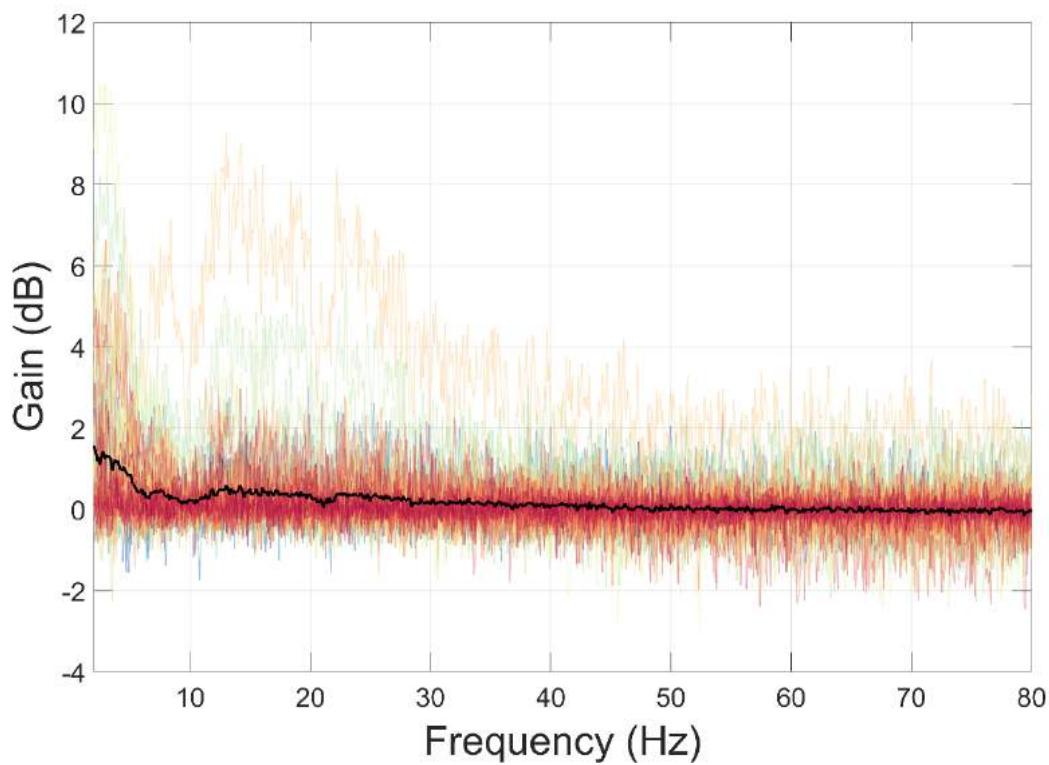

**Fig. 16.** The amount of interference suppression following ICA was quantified using Equation 2. Individual channels are plotted in colour, and the mean over all channels is plotted in black.

*5.2.8  The gradual removal of interference*

To demonstrate how the previous pre-processing steps sequentially removed large magnetic field changes from interfering sources, we calculated the maximum field change over 1 s chunks of the continuous OPM data. As seen in Fig. 17A, compared with the raw data (red line, 7-70 pT change per 1 s chunk), each pre-processing step reduced the variation in magnetic fields to less than 1 pT per 1 s chunk following ICA (purple line).

In the frequency domain, we calculated mean PSD values across all MEG channels after subsequent pre-processing steps. As seen in Figure 17B (left, panel), a high-pass filter suppressed activity below 2 Hz (green line). Unlike synthetic gradiometry (blue line), HFC (black line) reduced PSD values below 5 Hz. This was followed by ICA, which reduced PSD values further (purple line). Figure 17B (right, panel), shows the same results, but for higher frequencies (5-80 Hz).

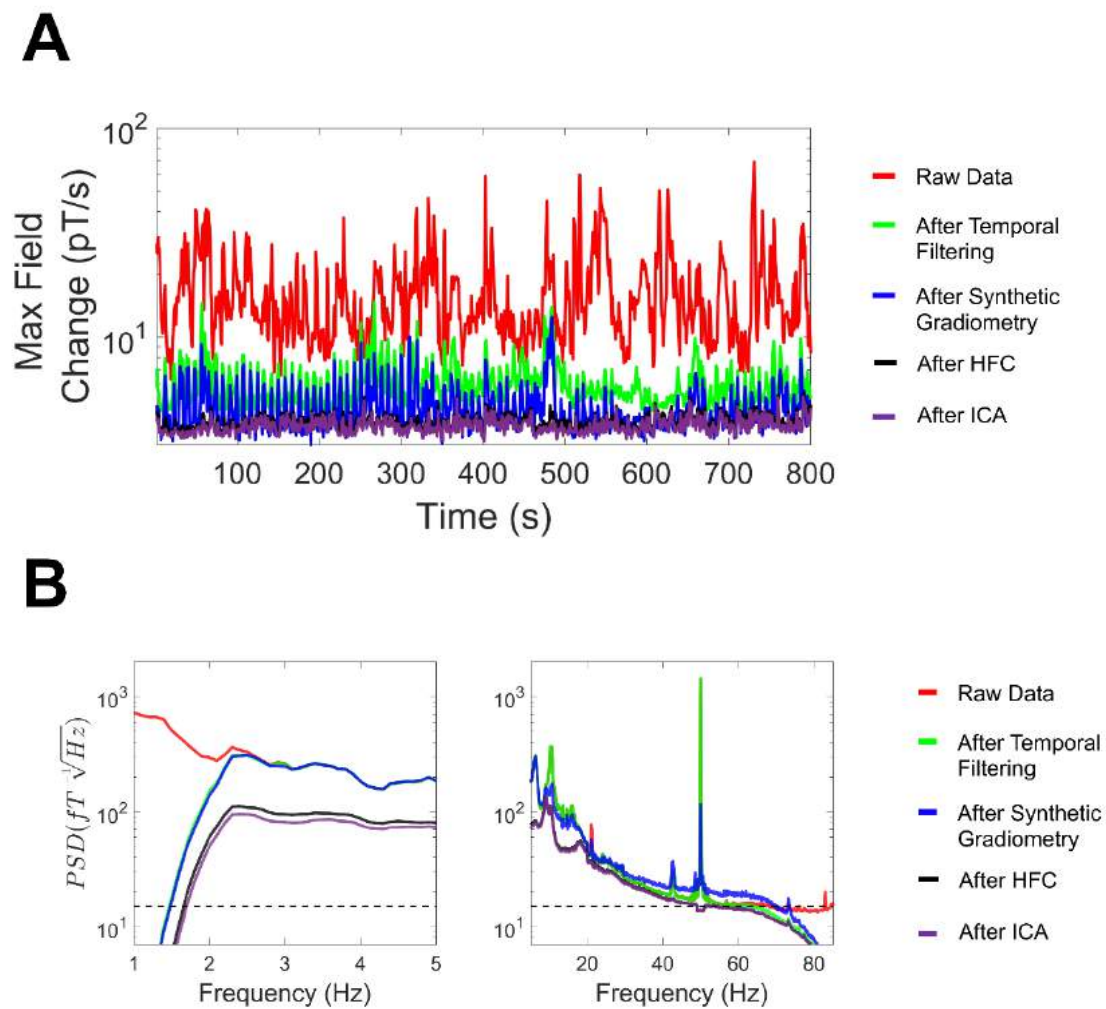

**Fig. 17.** (A) The maximum magnetic field change was calculated over 1 s chunks as each pre-processing stage was applied sequentially for the raw data, after temporal filtering, after synthetic gradiometry, after homogenous field correction (HFC), and after independent components analysis (ICA). (B) The power-spectral density (PSD) was calculated using 10 s-long windows, and averaged over channels, for the raw data and after each pre-processing step. The dotted line on the y-axis corresponds to 15 fT/√Hz.

*5.2.9 Sensor-level time-frequency analysis*

Pre-processed data were epoched into trials of 8 s (2 s pre-stimulus, 6 s post-stimulus onset), using a trigger sent when the fixation cross changed colour from white to red, indicating that the participant should start finger-tapping (OPM channel *NI-TRIG*). After 2.5 s the red cross changed back to white, indicating that the participant should stop finger-tapping.

Sensor-level time-frequency representations (TFRs) were calculated using a single Hanning taper between frequencies of 1–41 Hz in steps of 2 Hz (function: `ft_freqanalysis`). The entire 8 s epoch was used, with a sliding window of 500 ms, but the first and last 500 ms of each trial were discarded to avoid edge artefacts. All analyses were computed on single trials and subsequently averaged, and therefore TFRs contain both phase-locked (evoked) and non phase-locked (induced) information.

Human movement, including finger-tapping, involves the modulation of beta-band power in sensorimotor regions (Barratt et al., 2018; Cheyne, 2013), with decreases in power during movement, followed by increases (above baseline) following movement cessation (Neuper & Pfurtscheller, 2001). These are known as the event-related beta desynchronisation (ERBD) and post movement beta rebound (PMBR) respectively. We therefore focussed our TFR analysis on the beta (13-30 Hz) band. Field-maps were plotted to illustrate changes in beta power during finger tapping (0-2.5 s) and following cessation of finger tapping (2.5-4 s). The data were baseline-corrected using a baseline period of 1.5 s before stimulus onset and converted to dB. Only sensors oriented radially to the head were plotted, the tangential components being more difficult to visualise. Results show a robust ERBD (Fig. 18A, left panel), centred on left temporal/central sensors. This was followed by a PMBR from 2.5-4 s, again centred on left temporal/central sensors following movement cessation (Fig. 18A, right panel). The left-hemisphere bias in beta power modulation is consistent with the right-hand finger-tapping task performed by the participant, sensorimotor cortex being organised contralaterally.

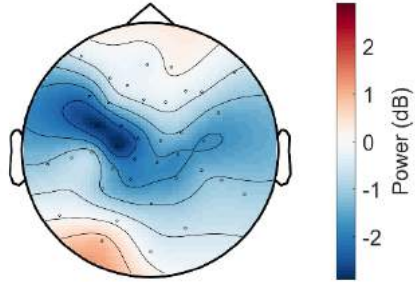
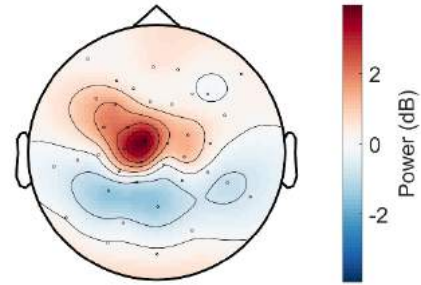
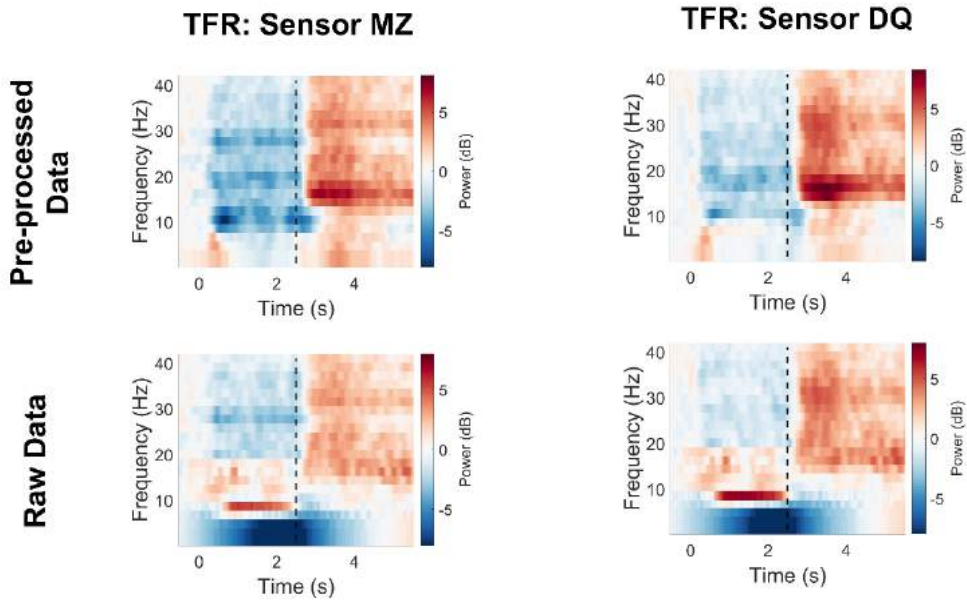

**Fig. 18.** (A). Fieldmaps were produced to show the topography of power changes in beta-band (13-30 Hz) power. On the left panel, event-related beta power is plotted during finger-tapping from 0-2.5 s. On the right panel, beta power is plotted following movement cessation from 2.5-4 s. The fieldmap only shows magnetic fields oriented radially to the head (the tangential components being more difficult to visualise). (B) Time-frequency representations (TFRs) were plotted for the sensors with the largest event-related desynchronisation (sensor MZ) and largest event-related synchronisation values (sensor DQ). The dotted black line at 2.5 s indicates the cessation of finger-tapping. On the top panel time-frequency power of the pre-processed data is plotted. On the bottom panel time-frequency power of the raw data is plotted.

To investigate these results in greater detail, TFRs were plotted for the sensors showing the maximum changes in ERBD (sensor *MZ*) and PMBR (sensor *DQ*). Both sensors were located over the left hemisphere approximately above sensorimotor cortex. As shown in Fig. 18B (top panel), for both sensors there is clear ERBD extending down to ~9 Hz, followed by PMBR at around 3 s after movement cessation centred on 15-35 Hz.

For illustration purposes, the TFR analysis was repeated using the raw data. Power was plotted for the sensors *MZ* and *DQ* (Fig. 18B, bottom panel). For both sensors, we can see that low-frequency power (under ~6 Hz) has an unusual smeared appearance. This is because none of the basis functions used for TFR analysis (a Hanning taper in this case) are exactly orthogonal to the high amplitude low-frequency interference present in the data. We can also observe how interference from 9-20 Hz is masking the ERBD from 0-2.5 s. Following movement cessation (the black dotted line), the PMBR is present in the raw data, however the power values are lower compared with the pre-processed data.

*5.2.10 Summary of the second tutorial*

In the second tutorial, we analysed OPM data from a stationary participant performing a finger-tapping task that is known to modulate beta-band (13-30 Hz) rhythms in sensorimotor cortex (Barratt et al., 2018; Cheyne, 2013). Despite lower movement-related artefacts below ~6 Hz, compared with the first tutorial, the raw data was still contaminated by interference across the frequency spectrum, including the beta-band. Without correction, these artefacts reduced the SNR and robustness of time-frequency analyses (see Fig. 18B, lower panel). The pipeline used to supress interference is summarised in Fig. 19. We first reduced power line noise and other sharp peaks in the power spectrum using spectral interpolation (Leske & Dalal, 2019). The data were then temporally filtered between 2-80 Hz. This was followed by synthetic gradiometry using the reference OPM data (Fife et al., 1999), which helped to reduce interference at ~10 Hz and 50 Hz power line noise. However, for sources of interference not measured by the reference sensors, especially lower frequency artefacts, synthetic gradiometry offered little benefit. In contrast, the spatial filtering technique HFC (Tierney et al., 2021a) helped to attenuate interference across the frequency spectrum. Finally, ICA was used to isolate and remove heart beat and eye-blink artefacts, helping to reduce interference below ~2 Hz. This pipeline, combining various different types of signal processing algorithms, increased the SNR of beta-band power modulations during the finger-tapping task at the sensor-level.

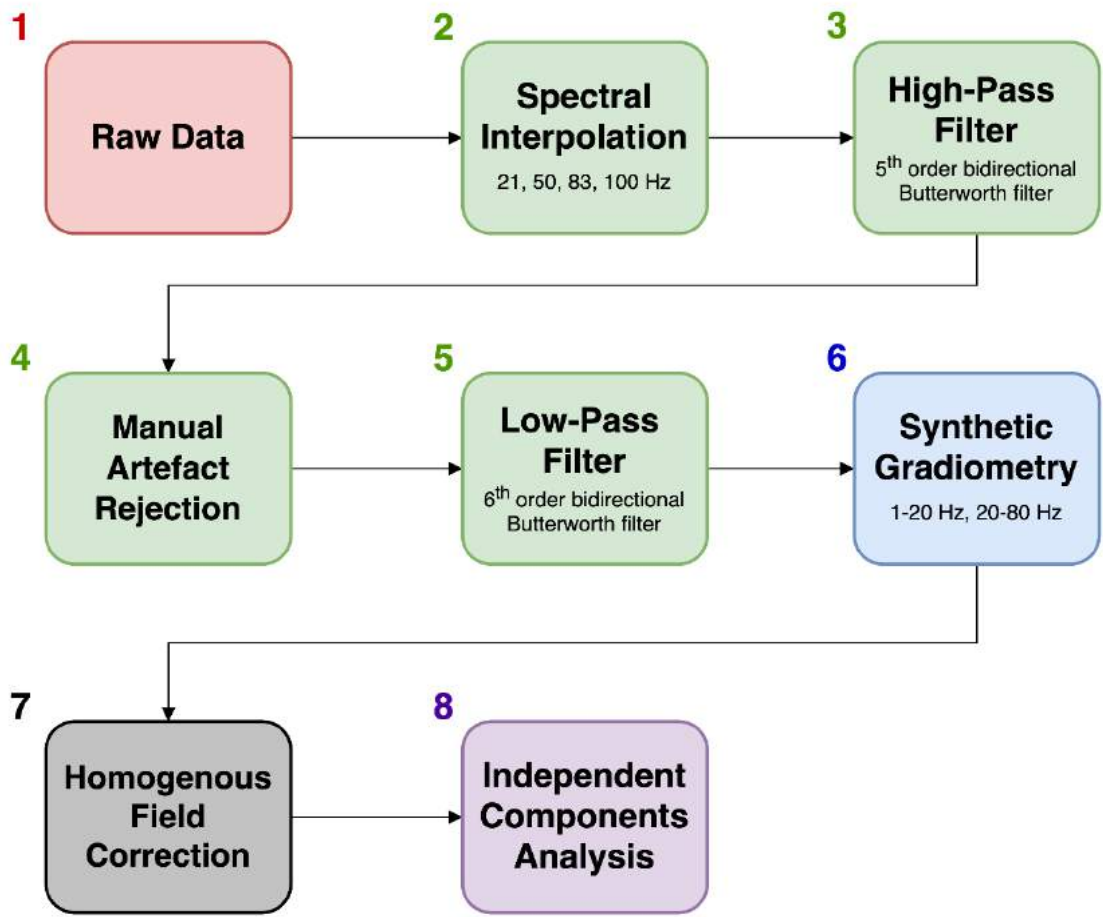

**Fig. 19.** Flow diagram demonstrating the order of the interference suppression steps taken in the second example tutorial. Note that colour coding aligns with the steps outlined in Fig. 17.

## 6. Conclusions

OPMs are opening up exciting new avenues for MEG research including paediatric measurements (Hill et al., 2019), and the adoption of more interactive, naturalistic paradigms involving movement (Holmes et al., 2021; Roberts et al., 2019; Seymour et al., 2021). OPMs are also likely to be used as a clinical tool in pre-epilepsy surgery planning (Feys et al., 2021; Mellor et al., 2021b; Vivekananda et al., 2020), and the study of mild traumatic brain injury (Allen et al., 2021). We have highlighted here the unique challenges facing OPMs in terms of interference suppression. The amount of noise measured by OPM-based MEG systems is likely to be far higher than conventional SQUID-MEG systems, especially in the context of mobile OPM experiments (Boto et al., 2017; Seymour et al., 2021), and naturalistic paradigms involving motion capture systems and virtual reality (Roberts et al., 2019). For this reason, methods aimed at reducing interfering magnetic fields are crucial for OPM-based MEG across a range of research and clinical contexts. In this article, a variety of different hardware solutions for interference suppression have been considered. We also outlined several signal processing approaches for attenuating interference, from a range of different sources, focussing on the practical application of these tools for OPM-based MEG data. The advent of multi-axis OPM recordings is likely to benefit these signal processing approaches further, especially those involving spatial filtering (Brookes et al., 2021). We also discussed how methods relying on the spatial oversampling of neuromagnetic fields, for example SSS (Taulu & Kajola, 2005), will only be applicable to OPM data once the channel count approaches that of conventional SQUID-MEG systems.

Both our tutorials demonstrated that while OPM data can contain high levels of noise, the careful application of signal processing tools can substantially reduce interference and increase the SNR of both evoked magnetic field and time-frequency analyses at the sensor- and source-levels. We encourage the reader to download the accompanying OPM data and run the tutorials for themselves in MATLAB. The signal processing pipelines (see Figs. 11, 19) could also be adapted for novel OPM-based MEG data across a variety of contexts. However, given the rapid development of OPM hardware within a variety of contrasting magnetic noise levels, it is too early at this stage to recommend a 'gold-standard' data analysis pipeline. Looking to the future, there is great scope for methods development in this space using novel hardware (e.g. Holmes et al., 2021) and/or signal processing techniques tailored for OPM data (Mellor et al., 2021a; Tierney et al., 2021a). This will be especially important where experimental questions of interest are focussed on lower frequencies, especially low delta (<2 Hz), which remains challenging to measure with OPMs during participant movement.

## Declaration of competing interest

This work was partly funded by a Wellcome Collaborative Award that involves a collaboration agreement with QuSpin Inc.

## Data and code availability

The data that support the findings of this study are available from Zenodo: https://doi.org/10.5281/zenodo.5539414 under an Attribution-ShareAlike 4.0 International license. Analysis code is openly available on GitHub:

https://github.com/FIL-OPMEG/tutorials_interference.

## Funding information

This research was supported by a Wellcome Principal Research Fellowship to E.A.M. (210567/Z/18/Z), a Wellcome Collaborative Award (203257/Z/16/Z), a Wellcome Centre Award (203147/Z/16/Z), the EPSRC-funded UCL Centre for Doctoral Training in Medical Imaging (EP/L016478/1), the Department of Health's NIHR-funded Biomedical Research Centre at University College London Hospitals, and EPSRC (EP/T001046/1) funding from the Quantum Technology hub in sensing and timing (sub-award QTPRF02).

This research was funded in whole, or in part, by Wellcome (Grant numbers: 210567/Z/18/Z; 203257/Z/16/Z; 203147/Z/16/Z). For the purpose of Open Access, the authors have applied a CC BY public copyright licence to any Author Accepted Manuscript version arising from this submission.

## CRediT authorship contribution statement

**Robert A. Seymour:** Conceptualisation, Methodology, Software, Investigation, Formal Analysis, Writing – Original Draft; **Nicholas Alexander:** Conceptualisation, Methodology, Software, Investigation, Writing – Review and Editing; **Stephanie Mellor:** Resources, Writing – Review and Editing; **George C. O'Neill:** Software, Resources, Writing – Review and Editing; **Tim M. Tierney:** Software, Resources, Writing – Review and Editing; **Gareth R. Barnes:** Conceptualisation, Writing – Review and Editing; **Eleanor A. Maguire:** Conceptualisation, Methodology, Supervision, Funding Acquisition; Writing – Review and Editing.

## Acknowledgements

Thanks to Vladimir Litvak and Ashwini Oswal for valuable discussions, David Bradbury and Clive Negus for imaging support, Vishal Shah at QuSpin Inc. and David Woolgar at Magnetic Shields Ltd. for technical assistance, and Mark Lim at Chalk Studios for help with scanner-cast design and construction.